\documentclass[aps,prb,twocolumn,showpacs,preprintnumbers,amsmath,amssymb]{revtex4-1}

\usepackage{graphicx}
\usepackage{amsmath,amssymb}
\usepackage[mathlines]{lineno}
\usepackage{bm}
\usepackage[english]{babel}
\usepackage[cp1251]{inputenc}
\usepackage{color}

\begin{document}

%
%
%
%

\author{\firstname{Roman~S.} \surname{Savelev}$^1$}
\author{\firstname{Olga~N.} \surname{Sergaeva}$^{1,2}$}
\author{\firstname{Denis~G.} \surname{Baranov}$^{3,4}$}
\author{\firstname{Alexander~E.} \surname{Krasnok}$^{1,5}$}
\author{\firstname{Andrea} \surname{Al\`u}$^5$}
\affiliation{$^1$ITMO University, St.~Petersburg 197101, Russia}
\affiliation{$^2$Department of Mechanical$\And$Aerospace Engineering, University of Missouri, Columbia, MO, 65211, USA}
\affiliation{$^3$Department of Physics, Chalmers University of Technology, 412 96 Gothenburg, Sweden}
\affiliation{$^4$Moscow Institute of Physics and Technology, 9 Institutskiy per., Dolgoprudny 141700, Russia}
\affiliation{$^5$Department of Electrical and Computer Engineering, The University of Texas at Austin, Austin, Texas 78712, USA}

\title{Dynamically reconfigurable metal-semiconductor Yagi-Uda nanoantenna}


\begin{abstract}
We propose a novel type of tunable Yagi-Uda nanoantenna composed of metal-dielectric (Ag-Ge) core-shell nanoparticles. We show that, due to the combination of two types of resonances in each nanoparticle, such hybrid Yagi-Uda nanoantenna can operate in two different regimes. Besides the conventional nonresonant operation regime at low frequencies, characterized by highly directive emission in the forward direction, there is another one at higher frequencies caused by hybrid magneto-electric response of the core-shell nanoparticles. This regime is based on the excitation of the van Hove singularity, and emission in this regime is accompanied by high values of directivity and Purcell factor within the same narrow frequency range. Our analysis reveals the possibility of flexible dynamical tuning of the hybrid nanoantenna emission pattern via electron-hole plasma excitation by 100 femtosecond pump pulse with relatively low peak intensities $\sim$200 MW/cm$^2$.
\end{abstract}

\maketitle

\section{Introduction}

Optical nanoantennas enable enhancement and flexible manipulation of light on a scale much smaller than one free-space wavelength~\cite{AluPRL2008,HulstNP2011,AluBook}. Due to this property, they offer unique opportunities for applications such as optical communications~\cite{Alu2010}, photovoltaics~\cite{Atwater2010205}, non-classical light emission~\cite{Maksymov2010}, subwavelength light confinement and enhancement~\cite{Stockman2004}, sensing~\cite{Liu2011631}, and single-photon sources~\cite{Vuckovic}. A specific type of optical nanoantennas, the Yagi-Uda array, has recently received a widespread attention in the literature~\cite{Krasnok2013UFN}. Such nanoantennas consist of several small scatterers that operate similarly to their radio frequency analogues. Yagi-Uda nanoantennas composed of different scatterers, such as core-shell nanoparticles~\cite{EnghetaPRB2007}, plasmonic nanoparticles~\cite{Koenderink2009}, high-index dielectric nanoparticles~\cite{KrasnokOE2012} were recently studied. Regardless of the scatterer type, such nanoantennas are characterized by a high directivity over a relatively wide frequency range. An explanation for this behaviour, based on the chain eigenmode analysis, was given in Ref.~[\onlinecite{Koenderink2009}] for plasmonic nanoparticles, and it can be also applied to any type of scatterer with a dominant dipole response.

Usually, a Yagi-Uda nanoantenna consists of a reflector and one or several directors. However, as it was pointed out in Ref.~[\onlinecite{Liu2012ACS}], a Yagi-Uda nanoantenna composed of core-shell nanoparticles with both electric dipole (ED) and magnetic dipole (MD) resonant responses at the same frequency does not necessarily need a reflector particle, since backward scattering can be suppressed automatically, due to interference of electric and magnetic harmonics~\cite{KerkerJOSAb1983}. Despite such special feature, core-shell Yagi-Uda nanoantennas have not been studied in details by now.

\begin{figure}[!t]
\includegraphics[width=0.7\columnwidth]{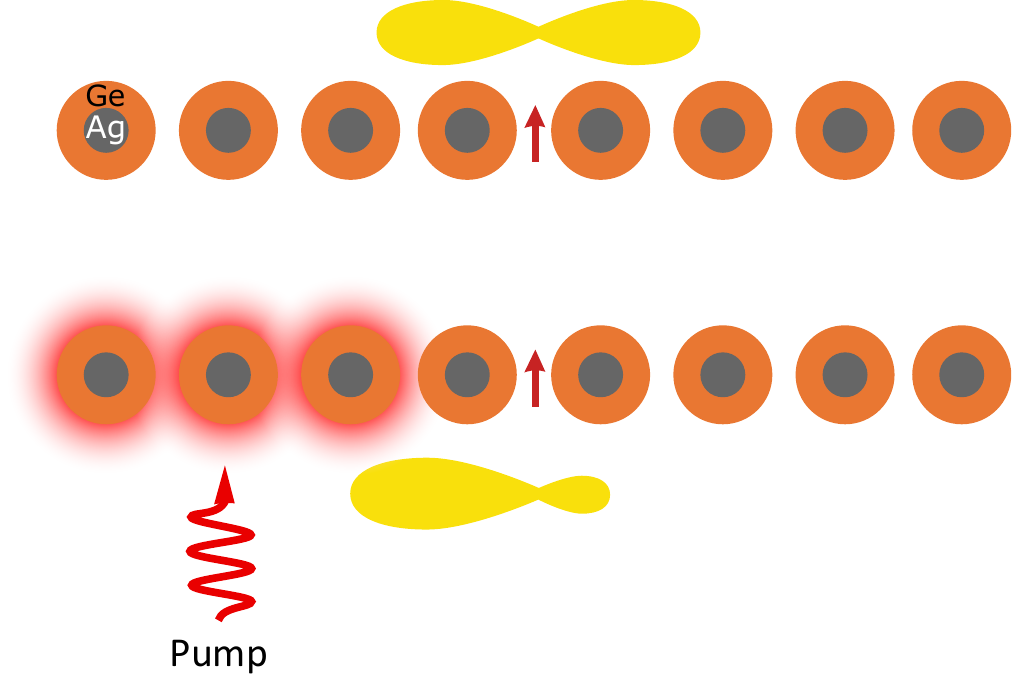}
\caption{Schematic of the tunable Yagi-Uda nanoantenna -- a periodic chain of metal-dielectric core-shell nanoparticles excited by a dipole emitter. The electric dipole emitter is placed in the center of the chain orthogonally to the chain axis. Generation of electron-hole plasma in the particles at the end of the chain shifts their resonances and dramatically modifies the emission pattern of the nanoantenna.}
\label{concept}
\end{figure}

In this paper, we theoretically investigate the capabilities of Yagi-Uda nanoantennas composed of Ag-Ge core-shell nanoparticles for tailoring the radiation of electric dipole emitters. We show that such a hybrid metal-dielectric nanoantenna can operate in two different regimes. In the first regime, the core-shell nanoantenna operates similarly to the well-known plasmonic and dielectric ones, and it is characterized by high directivity and low Purcell factor in a wide spectral range at low frequencies. Due to presence of both ED and MD resonances in a single core-shell nanosphere another operation regime is possible. In this regime, the effective excitation of a special class of dark magnetic dipole modes, associated with Van Hove singularity, in the chain of core-shell particles by an electric dipole source allows one to achieve both high directivity and high Purcell factor in a narrow (about 0.3\%) frequency range. We use this operation regime for efficient dynamical tuning of the Yagi-Uda nanoantenna properties via photoexcitation of electron-hole plasma (EHP) in dielectric shells by additional fs-laser signal pulses. Photoinduced generation of the plasma enables dramatic reconfiguration of the nanoantenna radiation pattern giving rise to strongly asymmetric emission directed along the chain (see Fig.~\ref{concept}), while maintaining high Purcell factor at the same time. 

\section{Core-shell nanoparticles as highly tunable nonlinear components}

As an element of the Yagi-Uda nanoantenna, we consider a core-shell nanoparticle with a silver (Ag) core and a germanium (Ge) shell. 
Bi-material core-shell nanoparticles find a broad range of applications in catalysis, nanoelectronics, and biophotonics~\cite{coreshell1, coreshell2}.
They also offer a flexible platform for manipulation of an electromagnetic radiation~\cite{DominguezNJoP2011,Liu2012ACS}. In contrast to single-material solid particles, the spectral position of the ED and MD resonances can be tuned almost independently over a wide range of frequencies by tuning their geometrical parameters, making such particles promising building blocks to realize functional nanophotonic devices.

The linear electromagnetic response of core-shell nanoparticles can be described using the generalized Mie theory for layered spherical particles~\cite{Kerker1951} (analytical expressions for scattering coefficients are given in Appendix A). In certain cases, the response of core-shell particles with a plasmonic core and a high-index dielectric shell can be well approximated as a combination of a plasmonic electric dipole (ED) resonant response and a magnetic dipole (MD) response due to the Mie-type resonance in the dielectric shell. A prominent example of such hybrid nanoparticles was studied in Refs.~[\onlinecite{Liu2012ACS},\onlinecite{DominguezNJoP2011}], where it was shown that for certain parameters the ED and MD resonances can completely overlap (with negligible higher order multipole resonances), giving rise to unidirectional scattering. In this work, we consider core-shell particles with parameters, close to those in Refs.~[\onlinecite{Liu2012ACS,DominguezNJoP2011}], and therefore we model them within the framework of the dipole approximation. In order to validate the main results we employ numerical simulations using CST Microwave Studio.

\begin{figure}[!t]
\center{\includegraphics[width=0.99\columnwidth]{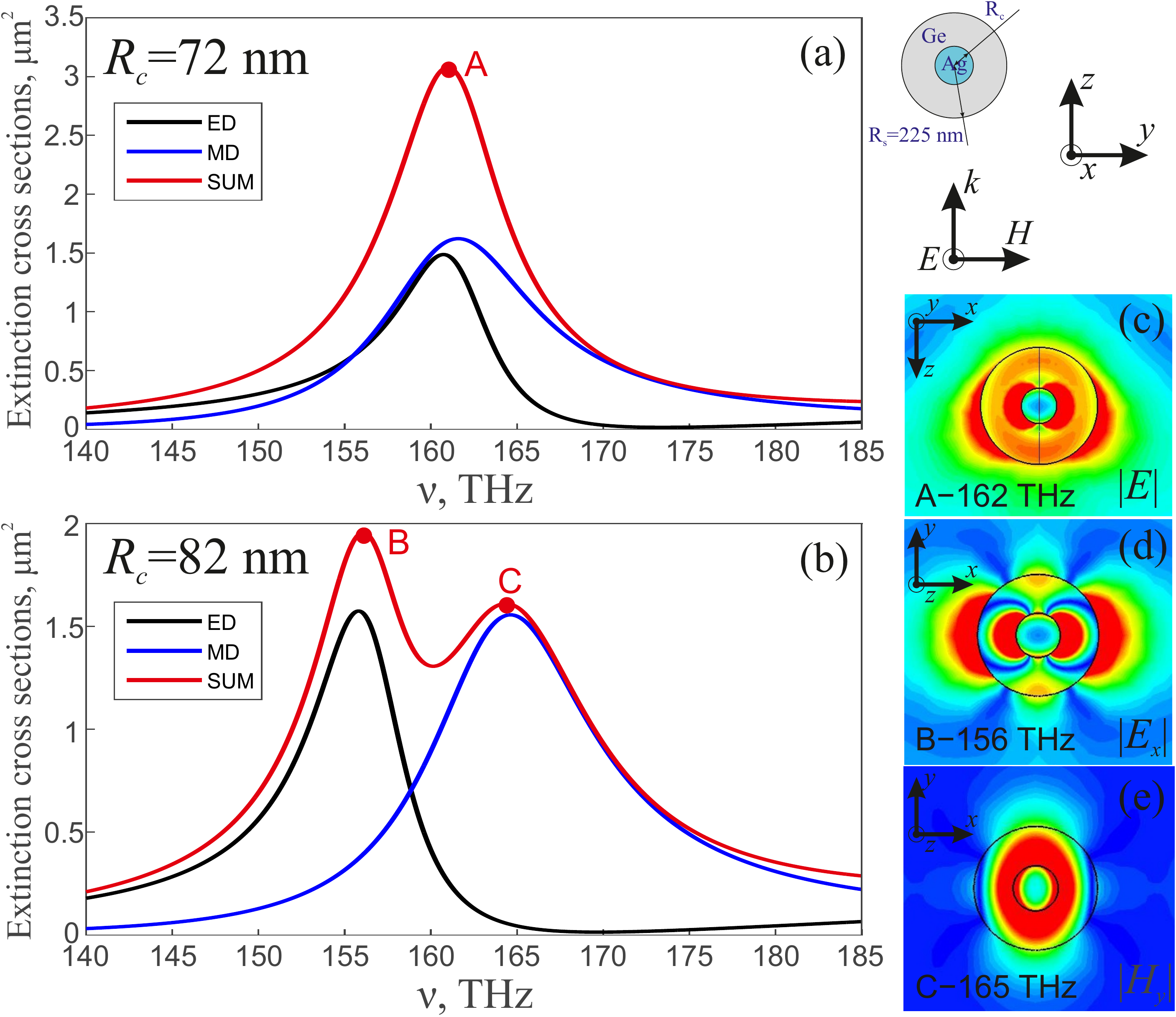}}
\caption{(a,b) Extinction cross-sections (red curves) for Ag-Ge core-shell particles with $R_s=225$~nm and (a) $R_c=72$~nm and (b) $R_c=82$~nm. Partial ED and MD contributions are shown with black and blue curves, respectively. (c-e) Field amplitude distributions at different frequencies: (c) point ``A'' in (a) (162~THz), (d) point ``B'' in (b) (156~THz), (e) point ``C'' in (b) (165~THz), indicating ED+MD, ED and MD resonances, respectively.}
\label{fig:RCS+fields}
\end{figure}

The Ag-Ge particles exhibit resonant dipole responses in the near infrared frequency range~\cite{DominguezNJoP2011}. For a radius of the silver core $R_c=72$~nm and a radius of the dielectric shell $R=225$~nm, an overlap of the ED and MD resonances of the Ag-Ge core-shell nanoparticle is achieved. We assume constant permittivity of germanium $\varepsilon_s=16.8$, and for the dispersion of silver core we employ a Drude model $\varepsilon_c \left(\nu \right) = 1 - \nu_p^2/(\nu^2 + {i \gamma \nu})$, with plasma frequency ${\nu_p} = 2180$ THz and collision frequency $\gamma = 4.93$ THz. At the frequency range of interest this model matches well the experimental data given in Ref.~[\onlinecite{JandC}]. The extinction cross sections of such a core-shell nanoparticle and their partial ED and MD contributions when excited by an $x$-polarized plane wave are shown in Figs.~\ref{fig:RCS+fields}(a,b) for $R_c=72$~nm and $R_c=82$~nm, respectively. The electric field distribution at the resonance frequency in Fig.~\ref{fig:RCS+fields}(a) (point ``A'') combines characteristic field distributions of both ED and MD resonances: namely, strongly enhanced electric field near the core due to the ED plasmonic resonance and slightly enhanced circularly distributed electric field due to the MD resonance in a high-index dielectric shell [see Fig.~\ref{fig:RCS+fields}(c)]. Increasing the core radius from 72~nm to 82~nm shifts the ED resonance frequency to $\approx$156~THz [point ``B'' in Fig.~\ref{fig:RCS+fields}(a)] and the MD resonance frequency to $\approx$165~THz [point ``C'' in Fig.~\ref{fig:RCS+fields}(a)] with corresponding field distributions plotted in Figs.~\ref{fig:RCS+fields}(d,e), respectively. This demonstrates the possibility of an efficient tuning of the two main resonances of the hybrid core-shell particle by simple changing their geometrical parameters.

\begin{figure*}[!t]
\center{\includegraphics[width=1\textwidth]{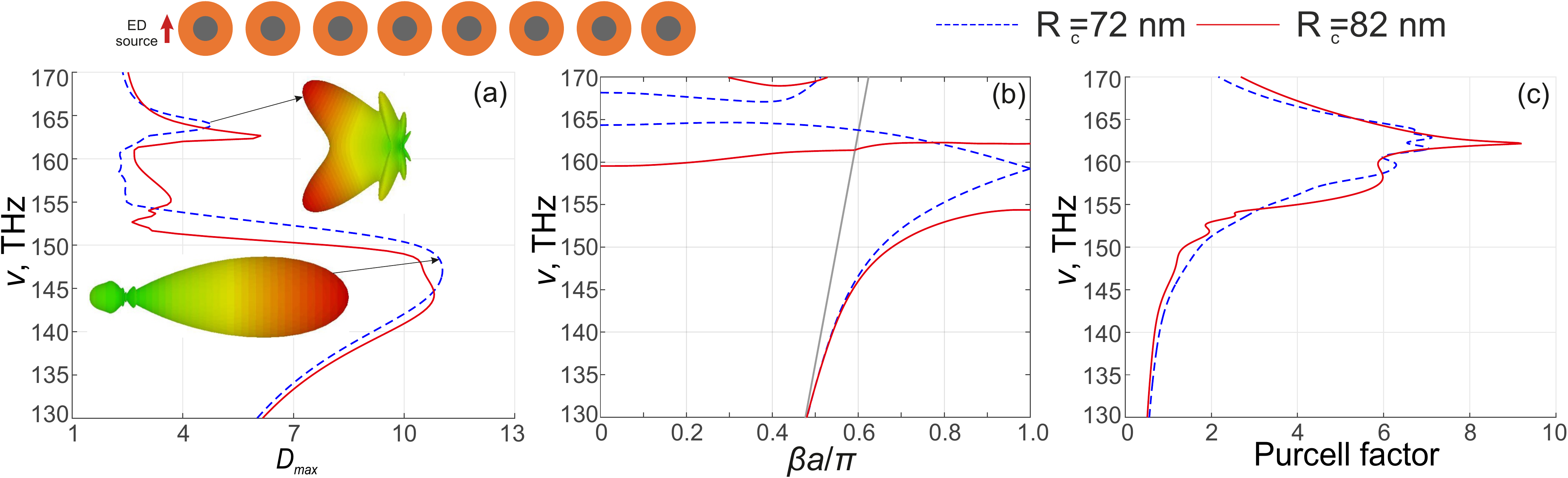}} \\
\caption{(a) Directivity and (c) Purcell factor for a chain of 8 core-shell nanoparticles with dipole source located on the axis of the chain at the half period from the center of the leftmost particle. (b) Dispersion diagram for the transversely polarized modes of an infinite chain of core-shell nanoparticles; $\beta a/\pi$ is the normalized Bloch wave number and $\nu$ is the frequency. Dashed blue (solid red) curves correspond to the following parameters: radius of Ag core $R_c=72$~nm ($R_c=82$~nm), radius of germanium shell $R_s=225$~nm and the period of the chain $a=550$~nm.}
\label{fig:72+82_side}
\end{figure*}

Beside the static ``geometrical'' tuning of core-shell particle properties, they can also be dynamically (and reversibly) tuned via photoexcitation of a dense electron-hole plasma in germanium shell. At normal conditions the conduction band of germanium is almost empty, however, optical absorption causes the electrons to fill the conduction band thus altering its permittivity and optical response~\cite{Sokolowski-Tinten2000, Gallant1982}. Therefore, one may employ the nonlinear response of the Ge shell associated with the EHP excitation for ultrafast tuning of light emission and scattering by the proposed core-shell particles. Recently, the photoexcitation of electron-hole plasma was employed for tuning of silicon nanoantennas in the IR and visible regions~\cite{Shcherbakov2015, Makarov2015, Baranov2016, Fischer2016}. In a recent work~\cite{BaranovLPR} this effect was utilized to achieve a beam steering in all-dielectric dimer nanoantennas. Here, we demonstrate that the same mechanism allows dramatic tuning of the emission pattern produced by a dipole source coupled with the proposed Yagi-Uda nanoantenna. Ge is a particularly good choice for plasma-induced nonlinear tuning of nanostuctures in the near-IR region, since it demonstrates a strong two-photon absorption ($\beta \approx 80$~cm/GW, Ref.~[\onlinecite{GeTPA}]) in this range, allowing to decrease the required pump intensity for significant detuning.

\section{Yagi-Uda core-shell nanoantennas}

It is known that the far fields produced by magnetic and electric dipoles induced by an impinging plane wave destructively interfere in the backward or forward direction if they have the same amplitude and are oscillating in or out of phase, respectively~\cite{KerkerJOSAb1983}. Since the backscattering is automatically suppressed in the case of particles with overlapping MD and ED resonances, the Yagi-Uda nanoantenna composed of such elements does not require a reflector~\cite{Liu2012ACS}. Therefore, we place a point electric dipole source (emulating a quantum emitter) oriented orthogonally to the axis of a chain of 8 core-shell nanospheres with $R_c=72$~nm and period $a=550$~nm. Frequency dependent directivity (see the details of calculations) for the considered nanoantenna is shown in Fig.~\ref{fig:72+82_side}(a). In order to understand the origin of the highly directive emission at low frequencies, following the analysis in Ref.~\onlinecite{Koenderink2009} we first calculate the dispersion properties of an infinite chain of core-shell nanospheres (see Appendix C for the details). In Fig.~\ref{fig:72+82_side}(b) we observe that the low frequency chain eigenmode is close to the light line at frequencies $\lesssim$ 150 THz (far from the particle resonances), which corresponds to the high directivity frequency range in Fig.~\ref{fig:72+82_side}(a). Besides the low frequency dispersion branch, due to the presence of two types of resonances there is also a second dispersion branch, which crosses the light line at $\approx$ 163~THz. For the considered parameters, ED and MD amplitudes of the eigenmodes are of the same order of magnitude for both branches and consequently there is a strong MD-ED interaction between neighboring particles. Due to this interaction, the second branch is characterized by a negative group velocity and, at the frequency near 163~THz, which corresponds to the observed peak in directivity, the nanoantenna radiates in the backward direction [see inset in Fig.~\ref{fig:72+82_side}(a)].

The conventional operation regime of Yagi-Uda nanoantennas at low frequencies is naturally nonresonant. This can be understood from the eigenmode analysis of the considered structure. While the infinite chain supports lossless modes (for lossless material and host medium) below the light line, eigenmodes of a finite chain are always radiative. Eigenfrequencies of the chain of 8 particles with the real part below 150~THz [corresponding to the first dispersion branch in Fig.~\ref{fig:72+82_side}(b)], have a very large imaginary part (see Supporting Information), and thus the low frequency operation of Yagi-Uda nanoantennas is broadband and nonresonant. Consequently, another important property of these nanoantennas, the Purcell factor (PF), is quite low in this regime~\cite{KrasnokOE2012}. We confirmed this property by calculating the PF within the framework of the dipole approximation (see Appendix B). Results of calculations are shown in Fig.~\ref{fig:72+82_side}(c): calculated values of PF at low frequencies do not exceed 1. The eigenfrequencies of the finite chain that correspond to the second dispersion branch, which is close to the resonances of the single particle, have small imaginary parts and therefore the PF is larger, but still $\lesssim$10.

\begin{figure}[!t]
	\center{\includegraphics[width=1\columnwidth]{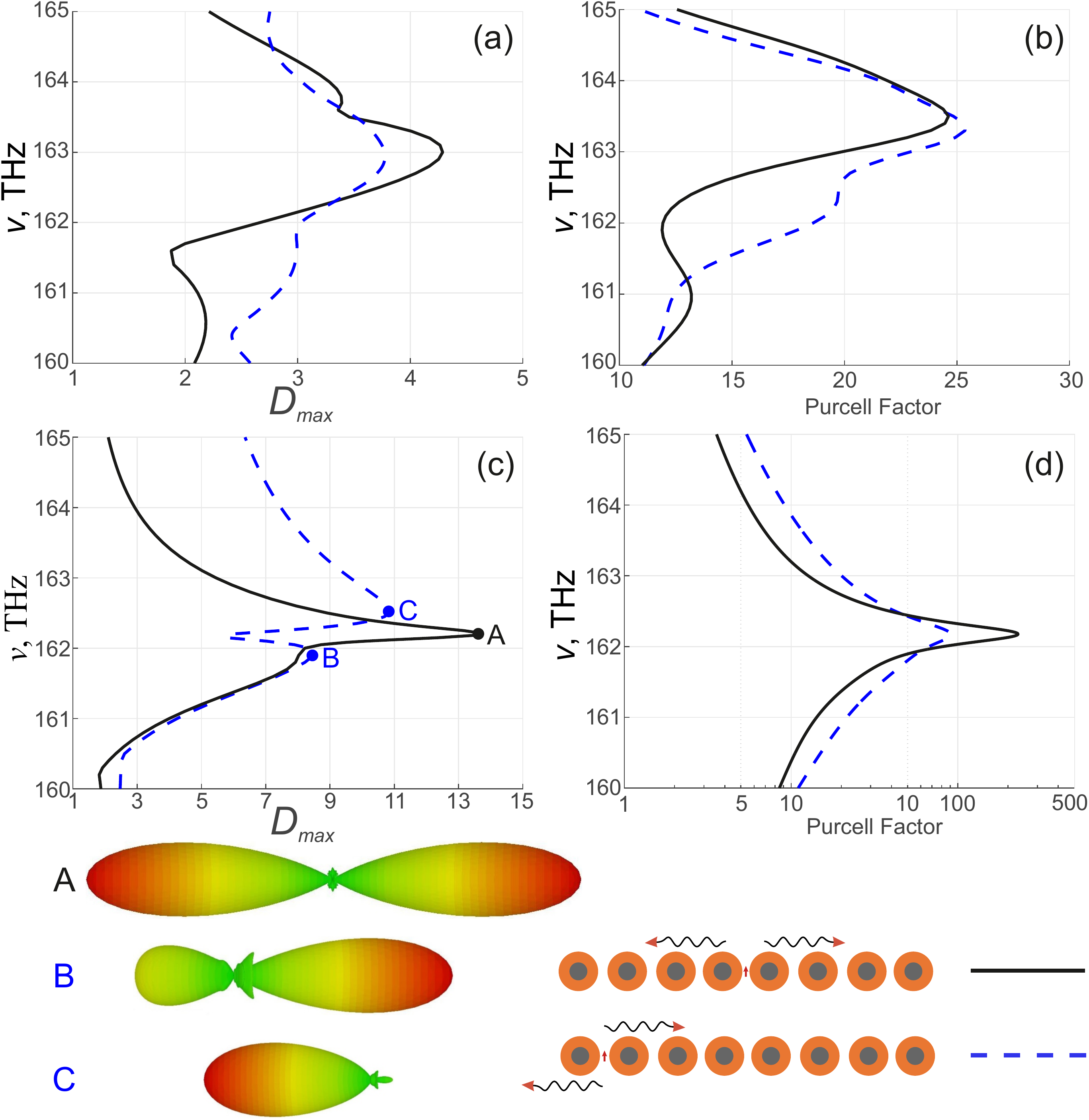}}
	\caption{(a,c) Directivity and (b,d) Purcell factor for the chain of 8 core-shell nanoparticles with dipole source located in the middle of the chain (solid black curves) and between the two leftmost particles (dashed blue curves). (a,b) correspond to the $R_c=72$~nm radius of Ag core and (c,d) --- to the $R_c=82$~nm; period of the chain is $a=550$~nm.}
	\label{fig:72+82_Mid_12}
\end{figure}

In order to increase the PF and, consequently, the antenna efficiency we employ the method of excitation of chain eigenmodes near the Van Hove singularity~\cite{KrasnokAPL2016}. This method relies on symmetry matching of the ED source oriented perpendicular to the chain axis and the staggered eigenmode of the chain of MDs, i.e., with MD moments in neighboring nanoparticles oscillating out-of-phase. If the source is placed outside the chain, the modes excitation is not very effective, which is the case demonstrated in Fig.~\ref{fig:72+82_side} for particles with two different radii of the core: in both cases the maximum of directivity and the PF are quite low in the resonant regime.

Therefore, we calculated the PF and directivity for two different positions of the source: in the center of the chain, and between the first and the second particles. For the case of $R_c=72$~nm, the directivity and PF in the narrow frequency range corresponding to the resonant operating regime are shown in Figs.~\ref{fig:72+82_Mid_12}(a,b). Despite the fact that for these source positions the chain eigenmodes are excited more effectively, we observe only a $\approx$4-fold increase in the PF, while the directivity decreases overall.

There are two reasons for the fact that the Purcell factor in this case is not very large. First, the dispersion of the array of hybrid MD-ED dipoles has an unusual behavior near the edge of the Brillouin zone. In the chain of particles with only one dominant dipole response (ED or MD) the group velocity tends to zero and, consequently, the density of photonic states diverges at the Van Hove singularity near the edge of the Brillouin zone (in the lossless case)~\cite{KrasnokAPL2016}. However, when the particles exhibit both MD and ED resonances at the same frequency, there is no gap between the dispersion branches (see Fig.~\ref{fig:72+82_side}(b) and Ref.~[\onlinecite{ShoreRS2012}] for the case of the particles with same permittivity and permeability). Since Purcell factors are proportional to the density of photonic states, very large values of PF cannot be achieved in the chain of core-shell nanoparticles with fully spectrally overlapping MD and ED resonances. Second, while there is a symmetry matching between the magnetic field of electric dipole sources placed between the particles and MD moments of the staggered chain eigenmode, the corresponding electric field and ED moments are not symmetrically matched. Since MD and ED moments in this case are of the same order, this also decreases the efficiency of excitation with electric dipole source.

A different situation is observed for core-shell particles with $R=82$~nm. In contrast to the case of $R=72$~nm, now the second branch is characterized by zero group velocity at the edge of the Brillouin zone, and an infinite local density of optical states [see Fig.~\ref{fig:72+82_side}(b)]. Moreover, the amplitudes of MD moments are substantially larger than those of ED moments for the second dispersion branch. Therefore, these eigenmodes can be effectively excited with ED sources, which allows achieving both high directivity and high rate of ED source emission in the same frequency range. Note that, since we are interested in the emission of ED sources, the MD resonance frequency should be larger than ED one. This is achieved because the first dispersion branch is characterized qualitatively by the same behavior for different relative position of resonances, and the frequency ranges of high directivity and high Purcell factor do not overlap. The second dispersion branch can instead be engineered in different ways.

For the case of $R=82$~nm, shown in Figs.~\ref{fig:72+82_Mid_12}(c,d), we observe that the electric dipole placed in the center of the chain, emits in both backward and forward directions along the chain with high directivity $\approx$14 (see insets in Fig.~\ref{fig:72+82_Mid_12}), while the PF for this configuration reaches high values up to 250 due to the high density of photonic states at the Van Hove singularity. Selective emission in a certain direction can be achieved by placing the source between the first and second particles. In this cases the PF is almost one order of magnitude lower, but the source emits predominantly in one direction at the operation frequencies (see insets in Fig.~\ref{fig:72+82_Mid_12}).

\section{Tuning of the nanoantenna radiation pattern}

\begin{figure}[!t]
\center{\includegraphics[width=1\columnwidth]{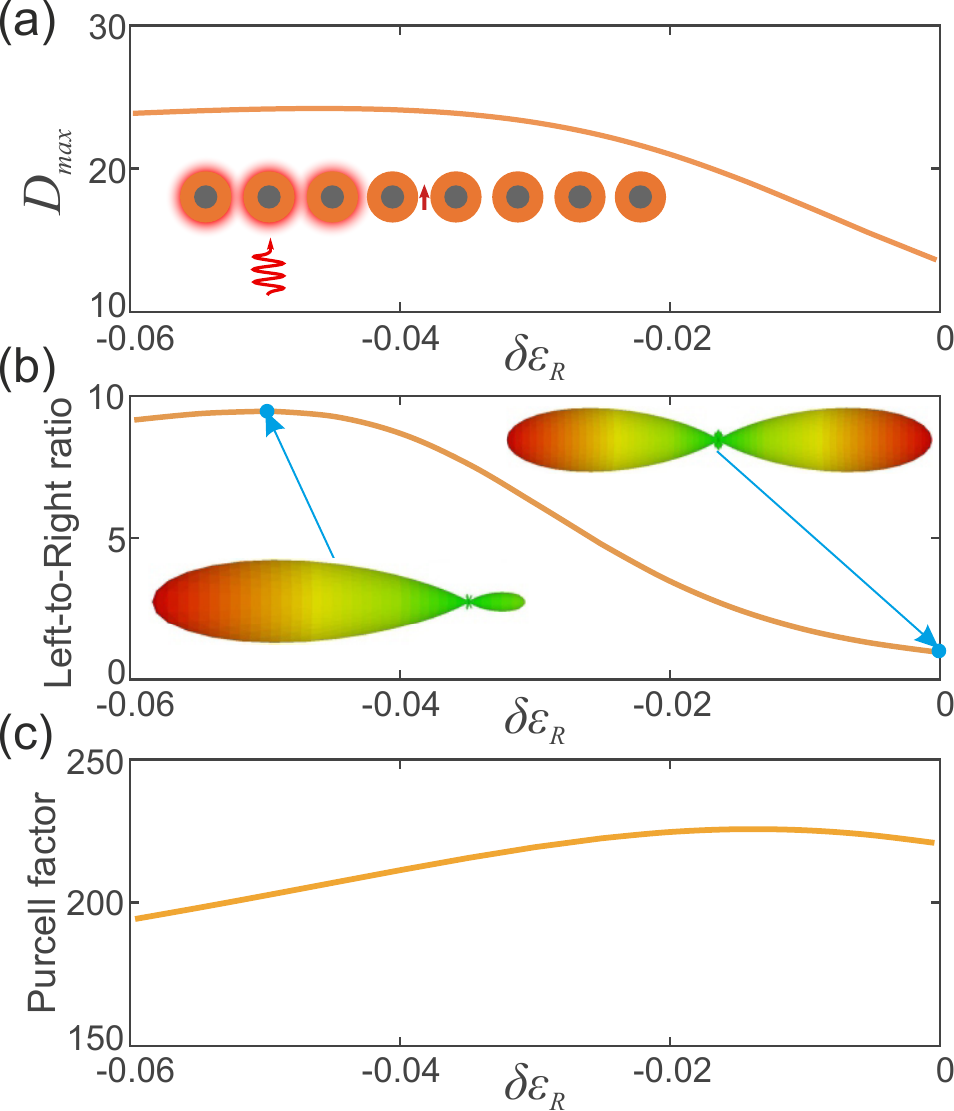}}
\caption{Tuning of the antenna emission via EHP generation. (a) Directivity, (b) Left-to-right ratio and (c) Purcell factor for the chain of 8 core-shell nanoparticles with a dipole source located in the middle of the chain versus the change in the $\mathrm{Re}(\varepsilon_p)$ of the three leftmost particles. Parameters of the chain are: $R_c=82$~nm, $R_s=225$~nm and $a=550$~nm. Operation frequency is 162.2~THz.}
\label{tuning}
\end{figure}

Now let us illustrate the capabilities of EHP photoexcitation to tailor the radiation pattern of the core-shell nanoantenna. The model to describe EHP-induced tuning of resonant spherical nanoparticles has been outlined in Refs.~\cite{Baranov2016, BaranovLPR}.
The permittivity of photoexcited Ge as a function of the EHP density $\rho_{eh}$ in the near-IR region is given by the expression~\cite{Gallant1982}
\begin{equation}
\varepsilon ^{{\text{exc}}}_{{\text{Ge}}}\left( {\omega ,{\rho _{{\text{eh}}}}} \right) = {\varepsilon _{{\text{Ge}}}}\left( \omega \right) + \delta {\varepsilon _R} + i\delta {\varepsilon _I},
\end{equation}
where 
\begin{equation}
\delta {\varepsilon _R} = - 4\pi {\rho _{eh}}{e^2}\left[ {\frac{1}{{m_e^*}}\frac{{\tau _e^2}}{{1 + {\omega ^2}\tau _e^2}} + \frac{1}{{m_h^*}}\frac{{\tau _h^2}}{{1 + {\omega ^2}\tau _h^2}}} \right]
\label{re}
\end{equation}
and
\begin{equation}
\delta {\varepsilon _I} = \frac{{4\pi {\rho _{eh}}{e^2}}}{\omega }\left[ {\frac{1}{{m_e^*}}\frac{{{\tau _e}}}{{1 + {\omega ^2}\tau _e^2}} + \frac{1}{{m_h^*}}\frac{{{\tau _h}}}{{1 + {\omega ^2}\tau _h^2}}} \right].
\label{im}
\end{equation}
The effective electron and hole masses for germanium are $m_e^*=0.12m_e$ and $m_h^*=0.23m_e$ with $m_e$ being the electron mass. Carriers collision times at room temperature can be taken as ${\tau _e} = {\tau _h} = 300$~fs~\cite{Gallant1982}.

The excitation of EHP in semiconductors is generally mediated by one- and two-photon absorption~\cite{Baranov2016}. In our situation, however, the two-photon process dominates, so that the rate equation governing the EHP dynamics may be written as
\begin{equation}
\frac{{d{\rho _{{\text{eh}}}}}}{{dt}} = - \Gamma {\rho _{{\text{eh}}}} + \frac{{{W_2}}}{{2\hbar \omega }}.
\label{rate}
\end{equation}
Here, $W_{2}$ is the volume-averaged two-photon absorption rate and $\Gamma$ is the EHP relaxation rate constant which depends on the EHP density. The absorption rate is written in the usual form ${W_2} = \frac{\omega }{{8\pi }} \left\langle {{{\left| {{\mathbf{\tilde E_{\rm in}}}} \right|}^4}} \right\rangle \operatorname{Im} {\chi ^{(3)}}$, where the angle brackets denote averaging over the semiconductor volume, and $\operatorname{Im} {\chi ^{(3)}} = \frac{ \varepsilon _{{\text{Ge}}} {c^2}} {{8\pi \omega }}\beta $ with $\beta$ being two-photon absorption coefficient ($\beta \approx 80$~cm/GW at 2.9~$\mu$m, Ref.~[\onlinecite{GeTPA}]). The EHP relaxation in Ge is mediated by Auger mechanism, $\Gamma=\Gamma_{\rm A} \rho_{eh}^{2}$ with $\Gamma_{\rm A}=7 \cdot 10^{ - 33}$~s$^{-1}$cm$^6$ (Ref.~[\onlinecite{GeAuger}]).

\begin{figure}[!t]
\includegraphics[width=0.9\columnwidth]{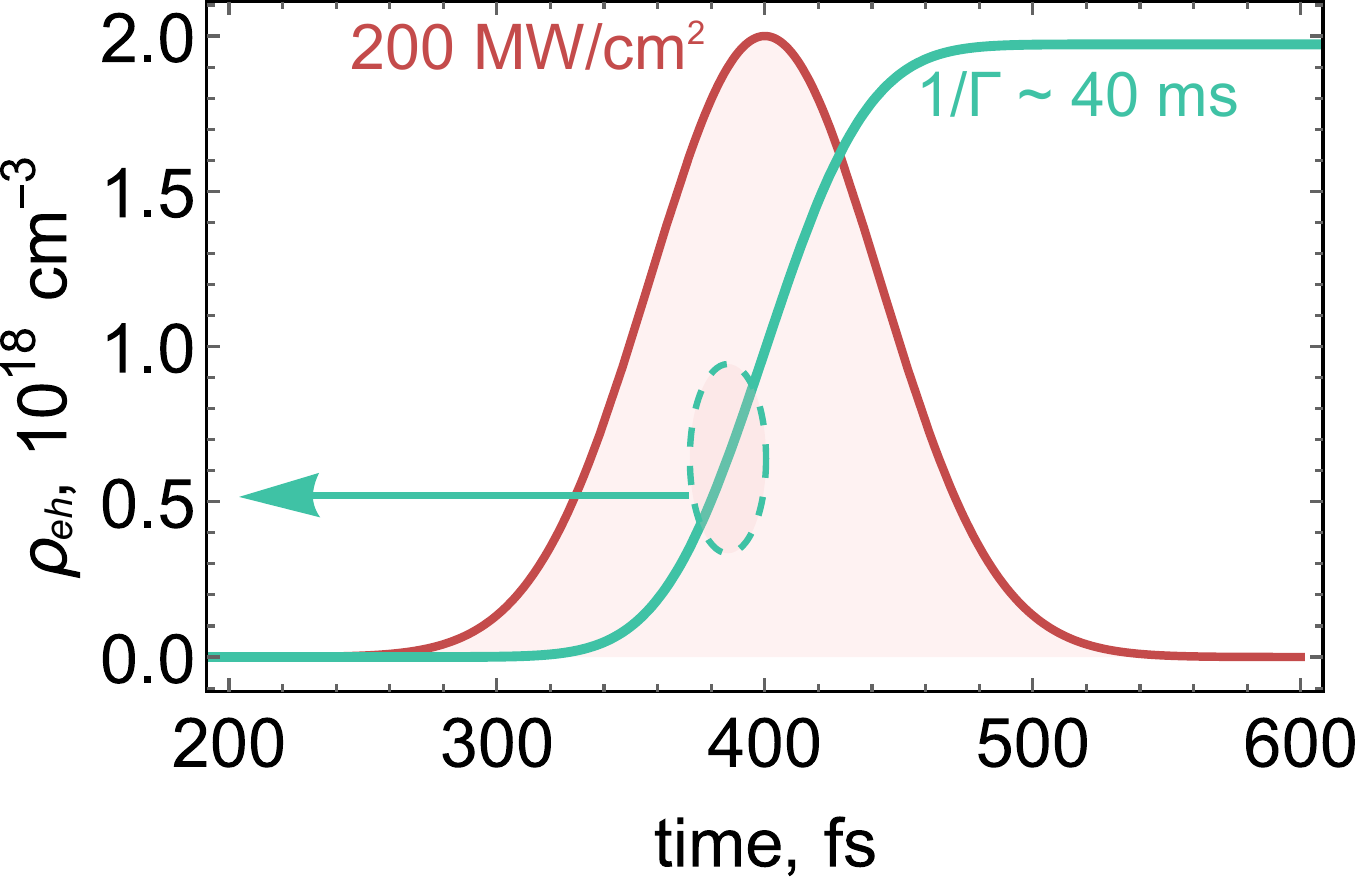}
\caption{EHP density in Ge shell of the core-shell nanoparticle illuminated with a 100 fs Gaussian pulse with 200 MW/cm$^2$ peak intensity. The shaded area represents the envelope of the pump pulse intensity. After the pulse action the EHP density decays exponentially with the time constant of about 40 ms.}
\label{pump}
\end{figure}

To demonstrate EHP-assisted tailoring of the emission pattern from the Yagi-Uda nanoantenna, we present in Fig.~\ref{tuning} the directivity, Left-to-Right ratio, and the Purcell factor of a point emitter located in the center of the chain as a function of real part of Ge permittivity change $\delta \varepsilon_R$. We assume that only the three leftmost particles are photoexcited, while the rest of the particles remains unexcited. The results clearly demonstrate that the radiation properties of the Yagi-Uda nanoantenna can be widely tuned via reasonable ($\sim-0.05$) modulation of the semiconductor permittivity. Less than 1\% decrease of Ge permittivity induces significant changes of the directivity [see Fig.~\ref{tuning}(a)], while retaining high values of Purcell factor [see Fig.~\ref{tuning}(c)]. But more interestingly, such permittivity modulation lifts the mirror symmetry of the antenna leading to large Left-to-Right ratio of nearly 10 [see Fig.~\ref{tuning}(b)], meaning that the emission of the source is radiated mostly from one end of the antenna.

In order to estimate the parameters of a pump pulse required for the given permittivity modulation, we use the rate equation~(\ref{rate}). The value of $\delta \varepsilon_R=-0.05$ is attained at the EHP density $\rho_{eh}\approx 2\times 10^{18}$~cm$^{-3}$ at the emitter wavelength $\lambda=1.6$~$\mu$m. At the same time, the increase of imaginary part of Ge permittivity at 1.6~$\mu$m given by Eq.~(\ref{im}) is negligible ($<10^{-4}$) at this level of EHP density, thus we may safely neglect its effect on the antenna emission. Numerical solution of Eq.~(\ref{rate}) indicates that a 100 fs Gaussian pulse with a peak intensity of 200 MW/cm$^2$ provides enough power to generate $2\times 10^{18}$~cm$^{-3}$ EHP, Fig.~\ref{pump}. The resulting intensity of the pump pulse is far below the damage threshold for plasmonic nanostructures.~\cite{Wurtz, Valev} After the pulse action, the Auger relaxation results in relatively slow recombination of the generated EHP with the characteristic time of approximately 40 ms, that should be sufficient for emission of a photon by typical quantum emitters.

\section*{Conclusion}

To conclude, we have studied the emission properties of a dipole source coupled to a Yagi-Uda nanoantenna composed of core-shell nanoparticles and demonstrated the possibility of dynamical tuning of the emission pattern via electron-hole plasma photoexcitation. The radiation pattern and emission rate of the nanoantenna fed by an electric dipole source strongly depend on the relative position of ED and MD resonance frequencies of a single core-shell particle. For most parameters, the hybrid core-shell nanoantenna operates as a conventional Yagi-Uda antenna without reflector, exhibiting highly directive emission over a wide frequency range. For certain parameters, an almost flat dispersion of the infinite chain of core-shell nanoparticles with an inherent high density of photonic states can be realized. In this regime, high values of Purcell factor and highly directive emission patterns can be achieved simultaneously.

Femtosecond optical pumping of the core-shell nanoparticles with moderate intensities accompanied by electron-hole plasma generation in Ge shells strongly affects the emission patterns of the nanoantenna allowing to direct the emission towards a chosen side of the chain. Our findings illustrate the potential of hybrid metal-dielectric nanoparticles and nanostructures, which can be employed for the development of directive single-photon emitters capable of ultrafast and reversible reconfiguration.


\appendix
\setcounter{secnumdepth}{0}
\section{Appendix A: Theoretical model}
To describe the electromagnetic response of an array of core-shell particles under consideration, we use the well-known coupled-dipole approximation (CDA) where each particle is characterized by electric dipole (ED) response (due to the plasmonic core) and magnetic dipole (MD) response (due to the high index dielectric shell). The analytical dipole model for a one-dimensional array of identical particles is formulated as follows~\cite{MulhollandLangmuir1994,MerchiersPRA2007,SavelevPRB2014,ShoreRS2012part1}:
\begin{align}
\begin{cases}
\mathbf{p}_n = \alpha_e(\omega)(\mathbf{E}^{\mathrm{(loc)}}_n+\mathbf{E}^{\mathrm{(ext)}}_n),\\
\mathbf{m}_n = \alpha_m(\omega)(\mathbf{H}^{\mathrm{(loc)}}_n+\mathbf{H}^{\mathrm{(ext)}}_n),
\label{eq:DM_freq_domain}
\end{cases}
\end{align}
where $\mathbf{m}_n$ and $\mathbf{p}_n$ are magnetic and electric dipole moments induced in the $n$th particle ($\propto -i\omega t$), $\alpha_m$ and $\alpha_e$ are the magnetic and electric polarizabilities; electric $\mathbf{E}^{\mathrm{(ext)}}_n$ and magnetic $\mathbf{H}^{\mathrm{(ext)}}_n$ fields at the position of the $n$th dipole are produced by external source; local electric $\mathbf{E}^{\mathrm{(loc)}}_n$ and magnetic $\mathbf{H}^{\mathrm{(loc)}}_n$ fields are produced by all other dipoles in the chain:
\begin{equation}
\begin{aligned}
\mathbf{E}^{\mathrm{(loc)}}_n &= \sum\limits_{j \ne n}\left( \widehat{C}_{nj} \mathbf{p}_j - \widehat{G}_{nj} \mathbf{m}_j \right),\\
\mathbf{H}^{\mathrm{(loc)}}_n &= \sum\limits_{j \ne n} \left( \widehat{C}_{nj} \mathbf{m}_j + \widehat{G}_{nj} \mathbf{p}_j \right),
\label{eq:field_loc}
\end{aligned}
\end{equation}
 where $\widehat{C}_{nj} = A_{nj}\widehat{I} + B_{nj}(\widehat{\mathbf{r}}_{nj} \otimes \widehat{\mathbf{r}}_{nj})$, $\widehat{G}_{nj} = - D_{nj}\widehat{\mathbf{r}}_{nj} \times \widehat{I}$, $\otimes$ is a dyadic product, $\widehat{I}$ is the unit $3 \times 3$ tensor, $\widehat{\mathbf{r}}_{nj}$ is the unit vector in the direction from $n$th to $j$th dipole, and
\begin{equation}
\begin{aligned}
A_{nj} &= e^{i k_h R_{nj}} \left( \dfrac{k_h^2}{R_{nj}} - \dfrac{1}{R_{nj}^3}+\dfrac{i k_h}{R_{nj}^2} \right),\\
B_{nj} &= e^{i k_h R_{nj}} \left( -\dfrac{k_h^2}{R_{nj}} + \dfrac{3}{R_{nj}^3} - \dfrac{3 i k_h}{R_{nj}^2} \right),\\
D_{nj} &= e^{i k_h R_{nj}} \left( \dfrac{k_h^2}{R_{nj}} + \dfrac{i k_h}{R_{nj}^2} \right),
\label{eq:int_const}
\end{aligned}
\end{equation}
where $R_{nj}$ is the distance between the $n$th and $j$th dipoles, $\varepsilon_h$ is the permittivity of the host medium (in our calculations we take $\varepsilon_h=1$), $k_h=\sqrt{\varepsilon_h}\omega/c$ is the wavenumber in the host medium, $\omega=2\pi \nu$, $\nu$ is the frequency, and $c$ is the speed of light. The electric and magnetic polarizabilities are defined as $\alpha_e = i\dfrac{3\varepsilon_h a_1^{sc}}{2k_h^3}$, $\alpha_m = i\dfrac{3 b_1^{sc}}{2k_h^3}$, where the scattering coefficients $a_1$, $b_1$ can be expressed through the parameters of the layered particle~\cite{Kerker1951}:
\begin{widetext}
\begin{equation}
\begin{aligned}
a_1^{sc} = \dfrac{\psi(\rho_s)[\psi'(n_s\rho_s) - A\chi'(n_s\rho_s)] - n_s\psi'(\rho_s)[\psi(n_s\rho_s) - A\chi(n_s\rho_s)]}{\xi(\rho_s)[\psi'(n_s\rho_s) - A\chi'(n_s\rho_s)] - n_s\xi'(\rho_s)[\psi(n_s\rho_s) - A\chi(n_s\rho_s)]}, \\
b_1^{sc} = \dfrac{n_s\psi(\rho_s)[\psi'(n_s\rho_s) - B\chi'(n_s\rho_s)] - \psi'(\rho_s)[\psi(n_s\rho_s) - B\chi(n_s\rho_s)]}{n_s\xi(\rho_s)[\psi'(n_s\rho_s) - B\chi'(n_s\rho_s)] - \xi'(\rho_s)[\psi(n_s\rho_s) - B\chi(n_s\rho_s)]},
\label{eq:scat_coefs}
\end{aligned}
\end{equation}
\end{widetext}
where $A =  \dfrac{n_s\psi(n_s\rho_c)\psi'(n_c\rho_c) - n_c\psi'(n_s\rho_c)\psi(n_c\rho_c)}{n_s\xi(n_s\rho_c)\psi'(n_c\rho_c) - n_c\xi'(n_s\rho_c)\psi(n_c\rho_c)}$, $B =  \dfrac{n_s\psi(n_c\rho_c)\psi'(n_s\rho_c) - n_c\psi'(n_c\rho_c)\psi(n_s\rho_c)}{n_s\xi'(n_s\rho_c)\psi(n_c\rho_c) - n_c\xi(n_s\rho_c)\psi'(n_c\rho_c)}$;\\ $\psi(x) = xj_1(x)$, $\chi(x) = -xy_1(x)$, $\xi(x) = xh_1^{(1)}(x)$ are the Ricatti-Bessel functions of the first order, $h_1^{(1)}(x) = j_1(x) + iy_1(x)$, $j_1(x)$, $y_1(x)$ are the first-order Hankel function of the first class, first order spherical Bessel function of the first and second class, respectively; $n_c=\sqrt{\varepsilon_c/\varepsilon_h}$ and $n_s = \sqrt{\varepsilon_s/\varepsilon_h}$ are the relative refractive indices of the core and the shell, respectively, $\rho_c=k_hR_c$ and $\rho_s=k_hR_c$.\par

\section{Appendix B: Calculation of the main properties of nanoantenna}

Directivity of a nanoantenna is calculated within the framework of the dipole model as follows~\cite{NovotnyNP2010}:
\begin{equation}
\begin{aligned}
D_{\bf max} = \dfrac {4\pi \bf Max \left[ \mathit p (\theta, \varphi) \right]}{P_{rad}},
\label{eq:Dir}
\end{aligned}
\end{equation}
where $\bf  Max \left[ \mathit p (\theta, \varphi) \right]$ is the power transmitted in the direction of the main lobe, $P_{rad}$ is the total power radiated by a system into the far zone, i.e., $P_{rad} = \int p (\theta, \varphi) d \Omega$ is the integral of the angular distribution $p (\theta, \varphi)$ of the radiated power over the spherical surface, where $(\theta, \varphi)$ are the angular coordinates of the spherical coordinate system, and $d \Omega$ is the element of the solid angle. The Purcell factor $PF$ is calculated with the well-known formula~\cite{NovotnyPrinciplesNanoOptics}:
\begin{equation}
\label{Purcell}
{\rm PF}=1+\dfrac{3}{2k^{3}d^{2}}\mbox{Im}\left[ {\bf d}^{*}\cdot{\bf E}_s \right],
\end{equation}
where $k$ is the wave number, $\mathbf{d}$ is the dipole moment of the emitter, and $\mathbf{E}_s$ is the scattered electric field at the position of the source.

\section{Appendix C: Calculation of chain eigenmodes}
\begin{figure}[h]
\center{\includegraphics[width=1\columnwidth]{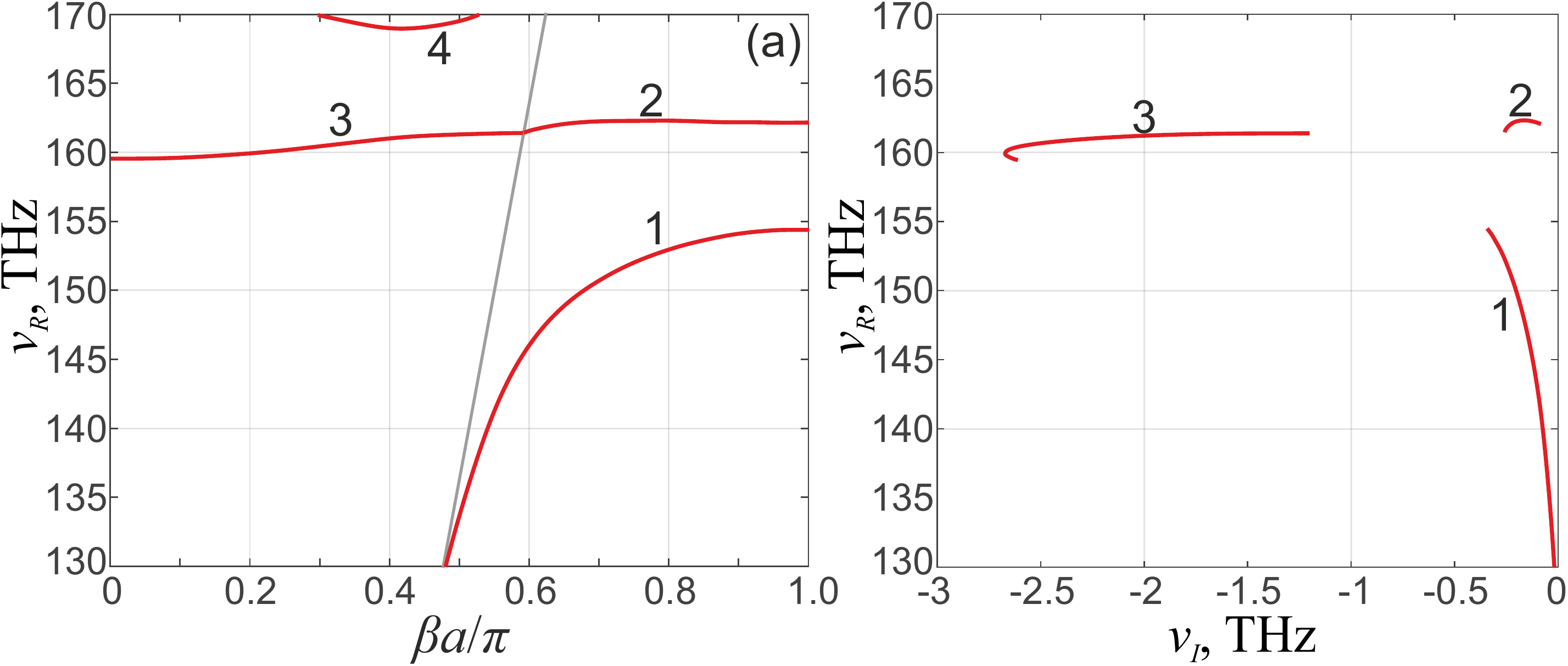}} \\
\caption{Eigenfrequencies of the infinite chain of core-shell particles with $R_c=82$\;nm, $R_s=225$\;nm placed with period $a=550$\;nm. (a) Real part of eigenfrequency $\nu_R$ vs. normalized Bloch wavenumber $\beta a/\pi$. (b) Real part of eigenfrequency $\nu_R$ vs. imaginary part of eigenfrequency $\nu_I$.}
\label{fig:EMs_inf}
\end{figure}

Eigenmodes of the infinite chain of core-shell nanoparticles are calculated from the analytical closed-form dispersion equations that can be obtained by replacing infinite sums in~(\ref{eq:field_loc}) with analytical functions~\cite{AluPRB2006}. In Fig.~3 in the main text only real parts of the complex eigenfrequencies are shown. In Fig.~\ref{fig:EMs_inf} we show both real and imaginary parts for the same parameters. The eigenmodes under light line have small non-zero imaginary parts due to losses in silver.\par

\begin{figure}[h]
\center{\includegraphics[width=1\columnwidth]{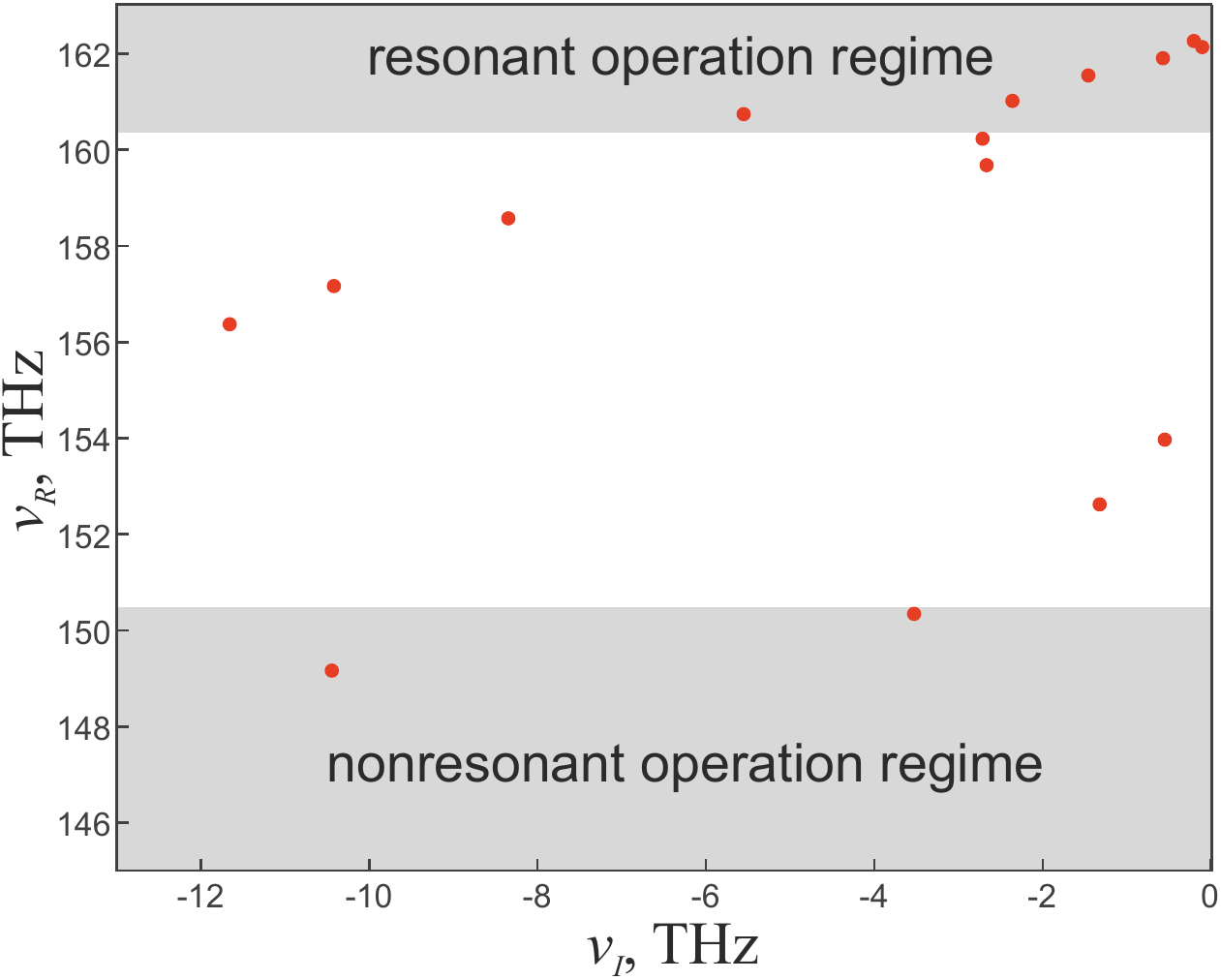}} \\
\caption{Complex eigenfrequencies ($\nu=\nu_R+i\nu_I$) of the finite chain of 8 core-shell particles with $R_c=82$\;nm, $R_s=225$\;nm placed with period $a=550$\;nm.}
\label{fig:EMs_fin}
\end{figure}

\begin{figure}[t!]
\center{\includegraphics[width=0.9\columnwidth]{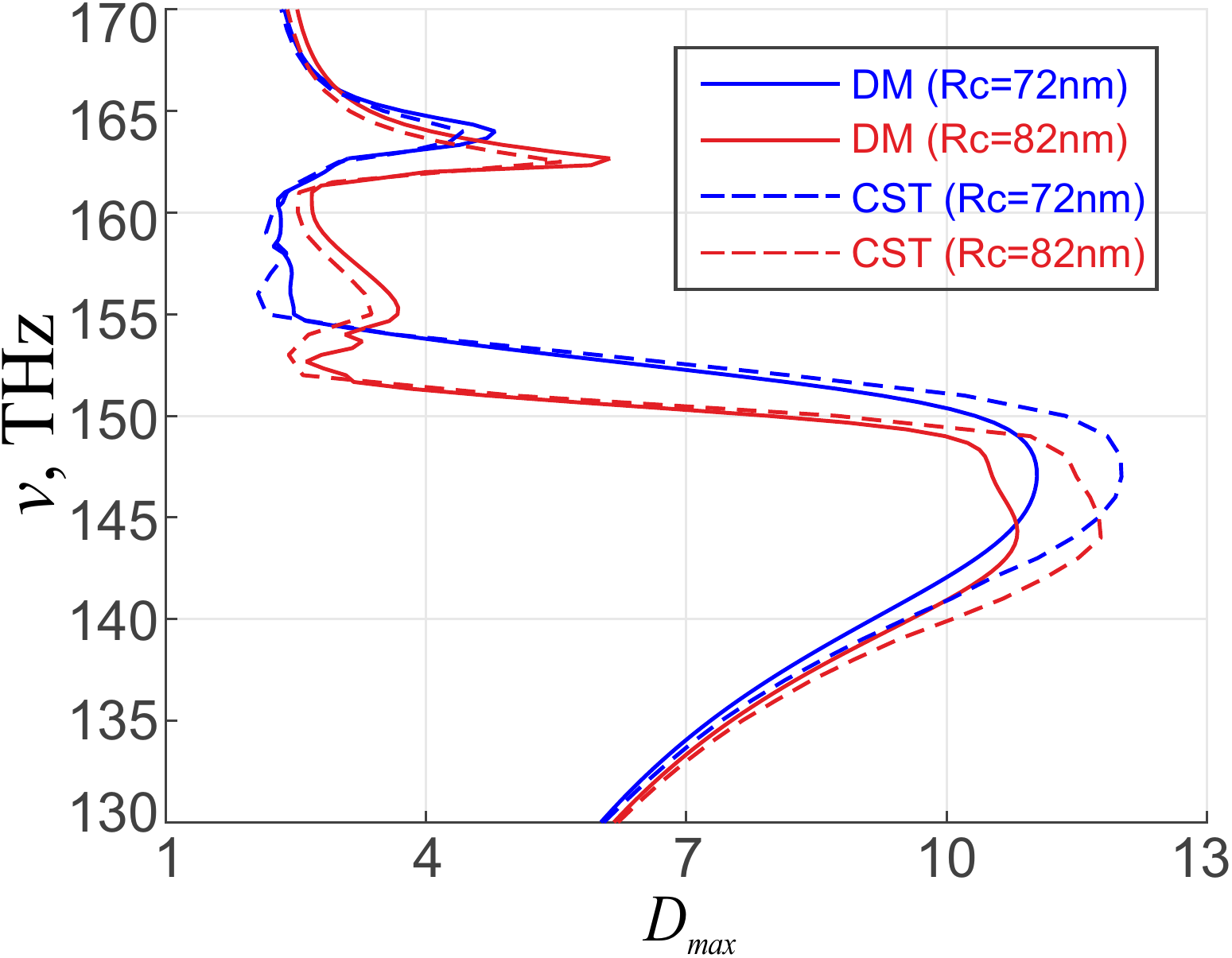}
\caption{Directivity for the chain of 8 core-shell nanoparticles with dipole source located on the axis of the chain at the half period from the center of the leftmost particle. Blue (red) curves correspond to the following parameters: radius of Ag core $R_c=72$~nm ($R_c=82$~nm), radius of germanium shell $R_s=225$~nm and the period of the chain $a=550$~nm. Solid curves are obtained within the framework of dipole approximation, dashed curves are obtained via numerical simulation.}} \label{fig:CST+DM_fig3}
\end{figure}
\begin{figure}[t!]
\qquad\center{\includegraphics[width=0.9\columnwidth]{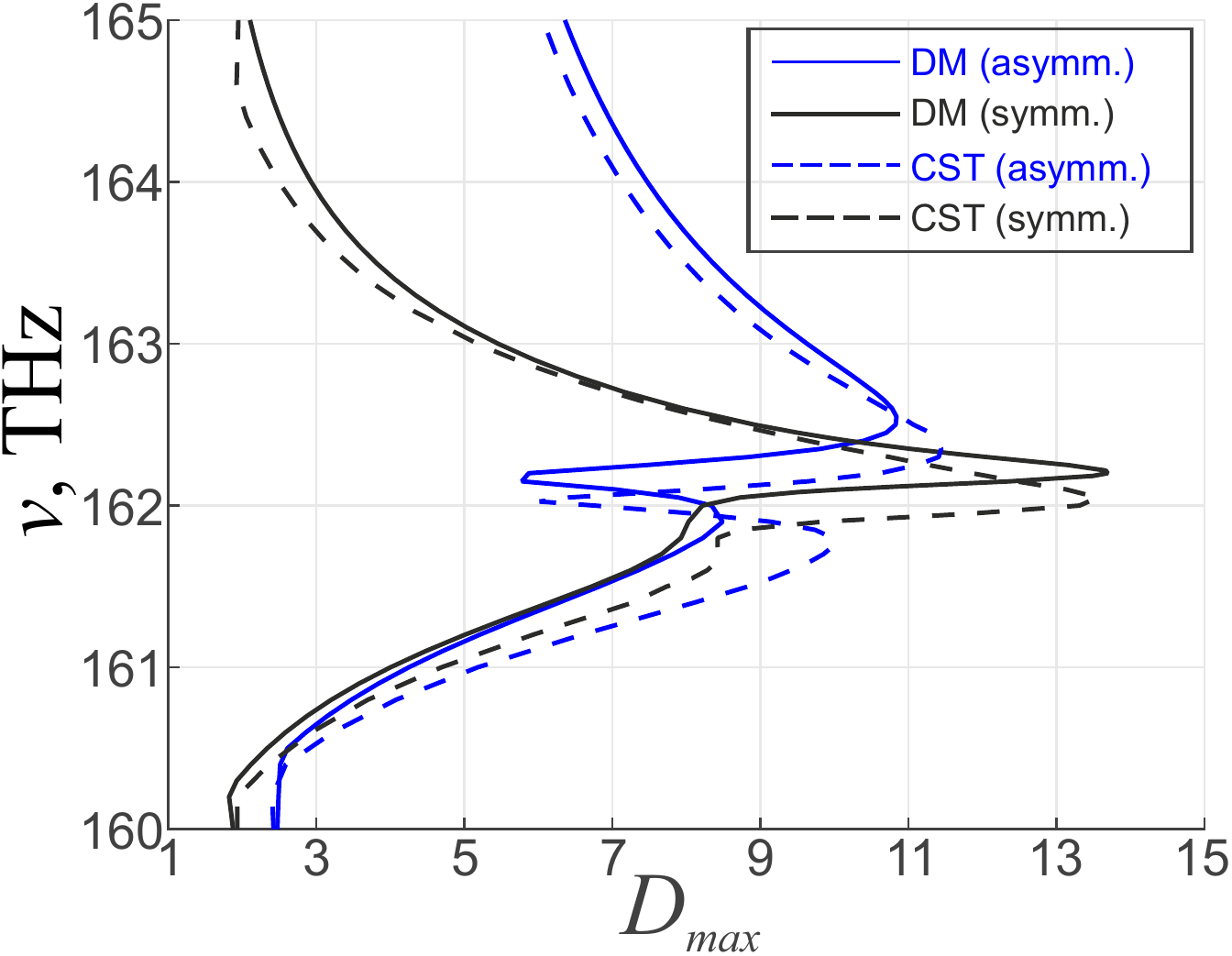}}
\caption{Directivity for the chain of 8 core-shell nanoparticles with dipole source located on the axis of the chain in the center of the chain (black curves) or between the first and the second particles (blue curves). Parameters of the structure: radius of Ag core $R_c=72$~nm ($R_c=82$~nm), radius of germanium shell $R_s=225$~nm and the period of the chain $a=550$~nm. Solid curves are obtained within the framework of dipole approximation, dashed curves are obtained via numerical simulation.} \label{fig:CST+DM_fig4}
\end{figure}

Eigenmodes of the finite chain of $N$ particles can be determined from the matrix equation~(\ref{eq:DM_freq_domain}) ($2N \times 2N$ in case of fixed trasnverse polarization) with zero right-hand side~\cite{WeberPRB2004}. The complex eigenfrequencies of the chain of 8 core-shell particles with $R_c=82$\;nm, $R_s=225$\;nm placed with period $a=550$\;nm are shown on the complex plane $[\mathrm{Re}(\omega),\mathrm{Im}(\omega)]$ in Fig.~\ref{fig:EMs_fin}. Modes with low real parts of eigenfrequencies, which correspond to the first dispersion branch ``1'' in Fig.~\ref{fig:EMs_inf}, have large imaginary parts, and, therefore, they are responsible for conventional nonresonant operation regime of the Yagi-Uda nanoantenna. Eigenfrequencies corresponding to the second dispersion branch ``2'' in Fig.~\ref{fig:EMs_inf} have smaller imaginary parts, and they are responsible for the resonant operation regime characterized with highly directive emission.

\section{Appendix D: Comparison with full-wave numerical simulations}

In Figs.~\ref{fig:CST+DM_fig3},\ref{fig:CST+DM_fig4} we show the direcitivity as a function of frequency for the same parameters, as in Figs.~\ref{fig:72+82_side}(a),\ref{fig:72+82_Mid_12}(c), respectively. Solid lines are calculated within the framework of the dipole approximation, dashed curves are directivities obtained via numerical simulation in CST Microwave Studio. We observe a good agreement between these two approaches.

%


\begin{thebibliography}{41}%
\makeatletter
\providecommand \@ifxundefined [1]{%
 \@ifx{#1\undefined}
}%
\providecommand \@ifnum [1]{%
 \ifnum #1\expandafter \@firstoftwo
 \else \expandafter \@secondoftwo
 \fi
}%
\providecommand \@ifx [1]{%
 \ifx #1\expandafter \@firstoftwo
 \else \expandafter \@secondoftwo
 \fi
}%
\providecommand \natexlab [1]{#1}%
\providecommand \enquote  [1]{``#1''}%
\providecommand \bibnamefont  [1]{#1}%
\providecommand \bibfnamefont [1]{#1}%
\providecommand \citenamefont [1]{#1}%
\providecommand \href@noop [0]{\@secondoftwo}%
\providecommand \href [0]{\begingroup \@sanitize@url \@href}%
\providecommand \@href[1]{\@@startlink{#1}\@@href}%
\providecommand \@@href[1]{\endgroup#1\@@endlink}%
\providecommand \@sanitize@url [0]{\catcode `\\12\catcode `\$12\catcode
  `\&12\catcode `\#12\catcode `\^12\catcode `\_12\catcode `\%12\relax}%
\providecommand \@@startlink[1]{}%
\providecommand \@@endlink[0]{}%
\providecommand \url  [0]{\begingroup\@sanitize@url \@url }%
\providecommand \@url [1]{\endgroup\@href {#1}{\urlprefix }}%
\providecommand \urlprefix  [0]{URL }%
\providecommand \Eprint [0]{\href }%
\providecommand \doibase [0]{http://dx.doi.org/}%
\providecommand \selectlanguage [0]{\@gobble}%
\providecommand \bibinfo  [0]{\@secondoftwo}%
\providecommand \bibfield  [0]{\@secondoftwo}%
\providecommand \translation [1]{[#1]}%
\providecommand \BibitemOpen [0]{}%
\providecommand \bibitemStop [0]{}%
\providecommand \bibitemNoStop [0]{.\EOS\space}%
\providecommand \EOS [0]{\spacefactor3000\relax}%
\providecommand \BibitemShut  [1]{\csname bibitem#1\endcsname}%
\let\auto@bib@innerbib\@empty
\bibitem [{\citenamefont {Al\`u}\ and\ \citenamefont
  {Engheta}(2008)}]{AluPRL2008}%
  \BibitemOpen
  \bibfield  {author} {\bibinfo {author} {\bibfnamefont {A.}~\bibnamefont
  {Al\`u}}\ and\ \bibinfo {author} {\bibfnamefont {N.}~\bibnamefont
  {Engheta}},\ }\href@noop {} {\bibfield  {journal} {\bibinfo  {journal} {Phys.
  Rev. Lett.}\ }\textbf {\bibinfo {volume} {101}},\ \bibinfo {pages} {043901}
  (\bibinfo {year} {2008})}\BibitemShut {NoStop}%
\bibitem [{\citenamefont {Novotny}\ and\ \citenamefont {van
  Hulst}(2011{\natexlab{a}})}]{HulstNP2011}%
  \BibitemOpen
  \bibfield  {author} {\bibinfo {author} {\bibfnamefont {L.}~\bibnamefont
  {Novotny}}\ and\ \bibinfo {author} {\bibfnamefont {N.}~\bibnamefont {van
  Hulst}},\ }\href@noop {} {\bibfield  {journal} {\bibinfo  {journal} {Nature
  Photon.}\ }\textbf {\bibinfo {volume} {5}},\ \bibinfo {pages} {83} (\bibinfo
  {year} {2011}{\natexlab{a}})}\BibitemShut {NoStop}%
\bibitem [{\citenamefont {Agio}\ and\ \citenamefont {Al\`u}(2013)}]{AluBook}%
  \BibitemOpen
  \bibfield  {author} {\bibinfo {author} {\bibfnamefont {M.}~\bibnamefont
  {Agio}}\ and\ \bibinfo {author} {\bibfnamefont {A.}~\bibnamefont {Al\`u}},\
  }\href@noop {} {\emph {\bibinfo {title} {Optical antennas}}}\ (\bibinfo
  {publisher} {Cambridge University Press, Cambridge},\ \bibinfo {year}
  {2013})\BibitemShut {NoStop}%
\bibitem [{\citenamefont {Al\`u}\ and\ \citenamefont
  {Engheta}(2010)}]{Alu2010}%
  \BibitemOpen
  \bibfield  {author} {\bibinfo {author} {\bibfnamefont {A.}~\bibnamefont
  {Al\`u}}\ and\ \bibinfo {author} {\bibfnamefont {N.}~\bibnamefont
  {Engheta}},\ }\href@noop {} {\bibfield  {journal} {\bibinfo  {journal} {Phys.
  Rev. Lett.}\ }\textbf {\bibinfo {volume} {104}},\ \bibinfo {pages} {213902}
  (\bibinfo {year} {2010})}\BibitemShut {NoStop}%
\bibitem [{\citenamefont {Atwater}\ and\ \citenamefont
  {Polman}(2010)}]{Atwater2010205}%
  \BibitemOpen
  \bibfield  {author} {\bibinfo {author} {\bibfnamefont {H.}~\bibnamefont
  {Atwater}}\ and\ \bibinfo {author} {\bibfnamefont {A.}~\bibnamefont
  {Polman}},\ }\href@noop {} {\bibfield  {journal} {\bibinfo  {journal} {Nat.
  Mater.}\ }\textbf {\bibinfo {volume} {9}},\ \bibinfo {pages} {205} (\bibinfo
  {year} {2010})}\BibitemShut {NoStop}%
\bibitem [{\citenamefont {Maksymov}\ \emph {et~al.}(2010)\citenamefont
  {Maksymov}, \citenamefont {Besbes}, \citenamefont {Hugonin}, \citenamefont
  {Yang}, \citenamefont {Beveratos}, \citenamefont {Sagnes}, \citenamefont
  {Robert-Philip},\ and\ \citenamefont {Lalanne}}]{Maksymov2010}%
  \BibitemOpen
  \bibfield  {author} {\bibinfo {author} {\bibfnamefont {I.~S.}\ \bibnamefont
  {Maksymov}}, \bibinfo {author} {\bibfnamefont {M.}~\bibnamefont {Besbes}},
  \bibinfo {author} {\bibfnamefont {J.~P.}\ \bibnamefont {Hugonin}}, \bibinfo
  {author} {\bibfnamefont {J.}~\bibnamefont {Yang}}, \bibinfo {author}
  {\bibfnamefont {A.}~\bibnamefont {Beveratos}}, \bibinfo {author}
  {\bibfnamefont {I.}~\bibnamefont {Sagnes}}, \bibinfo {author} {\bibfnamefont
  {I.}~\bibnamefont {Robert-Philip}}, \ and\ \bibinfo {author} {\bibfnamefont
  {P.}~\bibnamefont {Lalanne}},\ }\href@noop {} {\bibfield  {journal} {\bibinfo
   {journal} {Phys. Rev. Lett.}\ }\textbf {\bibinfo {volume} {105}},\ \bibinfo
  {pages} {180502} (\bibinfo {year} {2010})}\BibitemShut {NoStop}%
\bibitem [{\citenamefont {Stockman}(2004)}]{Stockman2004}%
  \BibitemOpen
  \bibfield  {author} {\bibinfo {author} {\bibfnamefont {M.}~\bibnamefont
  {Stockman}},\ }\href@noop {} {\bibfield  {journal} {\bibinfo  {journal}
  {Phys. Rev. Lett.}\ }\textbf {\bibinfo {volume} {93}},\ \bibinfo {pages}
  {137404} (\bibinfo {year} {2004})}\BibitemShut {NoStop}%
\bibitem [{\citenamefont {Liu}\ \emph {et~al.}(2011)\citenamefont {Liu},
  \citenamefont {Tang}, \citenamefont {Hentschel}, \citenamefont {Giessen},\
  and\ \citenamefont {Alivisatos}}]{Liu2011631}%
  \BibitemOpen
  \bibfield  {author} {\bibinfo {author} {\bibfnamefont {N.}~\bibnamefont
  {Liu}}, \bibinfo {author} {\bibfnamefont {M.}~\bibnamefont {Tang}}, \bibinfo
  {author} {\bibfnamefont {M.}~\bibnamefont {Hentschel}}, \bibinfo {author}
  {\bibfnamefont {H.}~\bibnamefont {Giessen}}, \ and\ \bibinfo {author}
  {\bibfnamefont {A.}~\bibnamefont {Alivisatos}},\ }\href@noop {} {\bibfield
  {journal} {\bibinfo  {journal} {Nat. Mater.}\ }\textbf {\bibinfo {volume}
  {10}},\ \bibinfo {pages} {631} (\bibinfo {year} {2011})}\BibitemShut
  {NoStop}%
\bibitem [{\citenamefont {Buckley}\ \emph {et~al.}(2012)\citenamefont
  {Buckley}, \citenamefont {Rivoire},\ and\ \citenamefont
  {Vuckovic}}]{Vuckovic}%
  \BibitemOpen
  \bibfield  {author} {\bibinfo {author} {\bibfnamefont {S.}~\bibnamefont
  {Buckley}}, \bibinfo {author} {\bibfnamefont {K.}~\bibnamefont {Rivoire}}, \
  and\ \bibinfo {author} {\bibfnamefont {J.}~\bibnamefont {Vuckovic}},\
  }\href@noop {} {\bibfield  {journal} {\bibinfo  {journal} {Rep. Prog. Phys.}\
  }\textbf {\bibinfo {volume} {75}},\ \bibinfo {pages} {126503} (\bibinfo
  {year} {2012})}\BibitemShut {NoStop}%
\bibitem [{\citenamefont {Krasnok}\ \emph {et~al.}(2013)\citenamefont
  {Krasnok}, \citenamefont {Maksymov}, \citenamefont {Denisyuk}, \citenamefont
  {Belov}, \citenamefont {Miroshnichenko}, \citenamefont {Simovski},\ and\
  \citenamefont {Kivshar}}]{Krasnok2013UFN}%
  \BibitemOpen
  \bibfield  {author} {\bibinfo {author} {\bibfnamefont {A.}~\bibnamefont
  {Krasnok}}, \bibinfo {author} {\bibfnamefont {I.}~\bibnamefont {Maksymov}},
  \bibinfo {author} {\bibfnamefont {A.}~\bibnamefont {Denisyuk}}, \bibinfo
  {author} {\bibfnamefont {P.}~\bibnamefont {Belov}}, \bibinfo {author}
  {\bibfnamefont {A.}~\bibnamefont {Miroshnichenko}}, \bibinfo {author}
  {\bibfnamefont {C.}~\bibnamefont {Simovski}}, \ and\ \bibinfo {author}
  {\bibfnamefont {Y.}~\bibnamefont {Kivshar}},\ }\href@noop {} {\bibfield
  {journal} {\bibinfo  {journal} {Phys. Usp.}\ }\textbf {\bibinfo {volume}
  {56}},\ \bibinfo {pages} {539} (\bibinfo {year} {2013})}\BibitemShut
  {NoStop}%
\bibitem [{\citenamefont {Li}\ \emph {et~al.}(2007)\citenamefont {Li},
  \citenamefont {Salandrino},\ and\ \citenamefont {Engheta}}]{EnghetaPRB2007}%
  \BibitemOpen
  \bibfield  {author} {\bibinfo {author} {\bibfnamefont {J.}~\bibnamefont
  {Li}}, \bibinfo {author} {\bibfnamefont {A.}~\bibnamefont {Salandrino}}, \
  and\ \bibinfo {author} {\bibfnamefont {N.}~\bibnamefont {Engheta}},\
  }\href@noop {} {\bibfield  {journal} {\bibinfo  {journal} {Phys. Rev. B}\
  }\textbf {\bibinfo {volume} {76}},\ \bibinfo {pages} {245403} (\bibinfo
  {year} {2007})}\BibitemShut {NoStop}%
\bibitem [{\citenamefont {Koenderink}(2009)}]{Koenderink2009}%
  \BibitemOpen
  \bibfield  {author} {\bibinfo {author} {\bibfnamefont {A.~F.}\ \bibnamefont
  {Koenderink}},\ }\href@noop {} {\bibfield  {journal} {\bibinfo  {journal}
  {Nano Lett.}\ }\textbf {\bibinfo {volume} {9}},\ \bibinfo {pages} {4228}
  (\bibinfo {year} {2009})}\BibitemShut {NoStop}%
\bibitem [{\citenamefont {Krasnok}\ \emph {et~al.}(2012)\citenamefont
  {Krasnok}, \citenamefont {Miroshnichenko}, \citenamefont {Belov},\ and\
  \citenamefont {Kivshar}}]{KrasnokOE2012}%
  \BibitemOpen
  \bibfield  {author} {\bibinfo {author} {\bibfnamefont {A.~E.}\ \bibnamefont
  {Krasnok}}, \bibinfo {author} {\bibfnamefont {A.~E.}\ \bibnamefont
  {Miroshnichenko}}, \bibinfo {author} {\bibfnamefont {P.~A.}\ \bibnamefont
  {Belov}}, \ and\ \bibinfo {author} {\bibfnamefont {Y.~S.}\ \bibnamefont
  {Kivshar}},\ }\href@noop {} {\bibfield  {journal} {\bibinfo  {journal} {Opt.
  Express}\ }\textbf {\bibinfo {volume} {20}},\ \bibinfo {pages} {20599}
  (\bibinfo {year} {2012})}\BibitemShut {NoStop}%
\bibitem [{\citenamefont {Liu}\ \emph {et~al.}(2012)\citenamefont {Liu},
  \citenamefont {Miroshnichenko}, \citenamefont {Neshev},\ and\ \citenamefont
  {Kivshar}}]{Liu2012ACS}%
  \BibitemOpen
  \bibfield  {author} {\bibinfo {author} {\bibfnamefont {W.}~\bibnamefont
  {Liu}}, \bibinfo {author} {\bibfnamefont {A.}~\bibnamefont {Miroshnichenko}},
  \bibinfo {author} {\bibfnamefont {D.}~\bibnamefont {Neshev}}, \ and\ \bibinfo
  {author} {\bibfnamefont {Y.}~\bibnamefont {Kivshar}},\ }\href {\doibase
  10.1021/nn301398a} {\bibfield  {journal} {\bibinfo  {journal} {ACS Nano}\
  }\textbf {\bibinfo {volume} {6}},\ \bibinfo {pages} {5489} (\bibinfo {year}
  {2012})}\BibitemShut {NoStop}%
\bibitem [{\citenamefont {Kerker}\ \emph {et~al.}(1983)\citenamefont {Kerker},
  \citenamefont {Wang},\ and\ \citenamefont {Giles}}]{KerkerJOSAb1983}%
  \BibitemOpen
  \bibfield  {author} {\bibinfo {author} {\bibfnamefont {M.}~\bibnamefont
  {Kerker}}, \bibinfo {author} {\bibfnamefont {D.-S.}\ \bibnamefont {Wang}}, \
  and\ \bibinfo {author} {\bibfnamefont {C.~L.}\ \bibnamefont {Giles}},\
  }\href@noop {} {\bibfield  {journal} {\bibinfo  {journal} {J. Opt. Soc. Am.}\
  }\textbf {\bibinfo {volume} {73}},\ \bibinfo {pages} {765} (\bibinfo {year}
  {1983})}\BibitemShut {NoStop}%
\bibitem [{\citenamefont {Zhong}\ and\ \citenamefont
  {Maye}(2001)}]{coreshell1}%
  \BibitemOpen
  \bibfield  {author} {\bibinfo {author} {\bibfnamefont {C.~J.}\ \bibnamefont
  {Zhong}}\ and\ \bibinfo {author} {\bibfnamefont {M.~M.}\ \bibnamefont
  {Maye}},\ }\href {\doibase
  10.1002/1521-4095(200110)13:19<1507::AID-ADMA1507>3.0.CO;2-#} {\bibfield
  {journal} {\bibinfo  {journal} {Adv. Mater.}\ }\textbf {\bibinfo {volume}
  {13}},\ \bibinfo {pages} {1507} (\bibinfo {year} {2001})}\BibitemShut
  {NoStop}%
\bibitem [{\citenamefont {Ghosh~Chaudhuri}\ and\ \citenamefont
  {Paria}(2012)}]{coreshell2}%
  \BibitemOpen
  \bibfield  {author} {\bibinfo {author} {\bibfnamefont {R.}~\bibnamefont
  {Ghosh~Chaudhuri}}\ and\ \bibinfo {author} {\bibfnamefont {S.}~\bibnamefont
  {Paria}},\ }\href@noop {} {\bibfield  {journal} {\bibinfo  {journal} {Chem.
  Rev.}\ }\textbf {\bibinfo {volume} {112}},\ \bibinfo {pages} {2373} (\bibinfo
  {year} {2012})}\BibitemShut {NoStop}%
\bibitem [{\citenamefont {Paniagua-Dom$\acute{\mathrm{i}}$nguez}\ \emph
  {et~al.}(2011)\citenamefont {Paniagua-Dom$\acute{\mathrm{i}}$nguez},
  \citenamefont {L$\acute{\mathrm{o}}$pez-Tejeira}, \citenamefont
  {Marqu$\acute{\mathrm{e}}$s},\ and\ \citenamefont
  {S$\acute{\mathrm{a}}$nchez-Gil}}]{DominguezNJoP2011}%
  \BibitemOpen
  \bibfield  {author} {\bibinfo {author} {\bibfnamefont {R.}~\bibnamefont
  {Paniagua-Dom$\acute{\mathrm{i}}$nguez}}, \bibinfo {author} {\bibfnamefont
  {F.}~\bibnamefont {L$\acute{\mathrm{o}}$pez-Tejeira}}, \bibinfo {author}
  {\bibfnamefont {R.}~\bibnamefont {Marqu$\acute{\mathrm{e}}$s}}, \ and\
  \bibinfo {author} {\bibfnamefont {J.~A.}\ \bibnamefont
  {S$\acute{\mathrm{a}}$nchez-Gil}},\ }\href@noop {} {\bibfield  {journal}
  {\bibinfo  {journal} {New J. Phys.}\ }\textbf {\bibinfo {volume} {13}},\
  \bibinfo {pages} {123017} (\bibinfo {year} {2011})}\BibitemShut {NoStop}%
\bibitem [{\citenamefont {Aden}\ and\ \citenamefont
  {Kerker}(1951)}]{Kerker1951}%
  \BibitemOpen
  \bibfield  {author} {\bibinfo {author} {\bibfnamefont {A.}~\bibnamefont
  {Aden}}\ and\ \bibinfo {author} {\bibfnamefont {M.}~\bibnamefont {Kerker}},\
  }\href {\doibase 10.1063/1.1699834} {\bibfield  {journal} {\bibinfo
  {journal} {J. Appl. Phys.}\ }\textbf {\bibinfo {volume} {22}},\ \bibinfo
  {pages} {1242} (\bibinfo {year} {1951})}\BibitemShut {NoStop}%
\bibitem [{\citenamefont {Johnson}\ and\ \citenamefont
  {Christy}(1972)}]{JandC}%
  \BibitemOpen
  \bibfield  {author} {\bibinfo {author} {\bibfnamefont {P.~B.}\ \bibnamefont
  {Johnson}}\ and\ \bibinfo {author} {\bibfnamefont {R.~W.}\ \bibnamefont
  {Christy}},\ }\href {\doibase 10.1103/PhysRevB.6.4370} {\bibfield  {journal}
  {\bibinfo  {journal} {Phys. Rev. B}\ }\textbf {\bibinfo {volume} {6}},\
  \bibinfo {pages} {4370} (\bibinfo {year} {1972})}\BibitemShut {NoStop}%
\bibitem [{\citenamefont {Sokolowski-Tinten}\ and\ \citenamefont {von~der
  Linde}(2000)}]{Sokolowski-Tinten2000}%
  \BibitemOpen
  \bibfield  {author} {\bibinfo {author} {\bibfnamefont {K.}~\bibnamefont
  {Sokolowski-Tinten}}\ and\ \bibinfo {author} {\bibfnamefont {D.}~\bibnamefont
  {von~der Linde}},\ }\href {\doibase 10.1103/PhysRevB.61.2643} {\bibfield
  {journal} {\bibinfo  {journal} {Phys. Rev. B}\ }\textbf {\bibinfo {volume}
  {61}},\ \bibinfo {pages} {2643} (\bibinfo {year} {2000})}\BibitemShut
  {NoStop}%
\bibitem [{\citenamefont {Gallant}\ and\ \citenamefont {{Van
  Driel}}(1982)}]{Gallant1982}%
  \BibitemOpen
  \bibfield  {author} {\bibinfo {author} {\bibfnamefont {M.~I.}\ \bibnamefont
  {Gallant}}\ and\ \bibinfo {author} {\bibfnamefont {H.~M.}\ \bibnamefont {{Van
  Driel}}},\ }\href {\doibase 10.1103/PhysRevB.26.2133} {\bibfield  {journal}
  {\bibinfo  {journal} {Phys. Rev. B}\ }\textbf {\bibinfo {volume} {26}},\
  \bibinfo {pages} {2133} (\bibinfo {year} {1982})}\BibitemShut {NoStop}%
\bibitem [{\citenamefont {Shcherbakov}\ \emph {et~al.}(2015)\citenamefont
  {Shcherbakov}, \citenamefont {Vabishchevich}, \citenamefont {Shorokhov},
  \citenamefont {Chong}, \citenamefont {Choi}, \citenamefont {Staude},
  \citenamefont {Miroshnichenko}, \citenamefont {Neshev}, \citenamefont
  {Fedyanin},\ and\ \citenamefont {Kivshar}}]{Shcherbakov2015}%
  \BibitemOpen
  \bibfield  {author} {\bibinfo {author} {\bibfnamefont {M.~R.}\ \bibnamefont
  {Shcherbakov}}, \bibinfo {author} {\bibfnamefont {P.~P.}\ \bibnamefont
  {Vabishchevich}}, \bibinfo {author} {\bibfnamefont {A.~S.}\ \bibnamefont
  {Shorokhov}}, \bibinfo {author} {\bibfnamefont {K.~E.}\ \bibnamefont
  {Chong}}, \bibinfo {author} {\bibfnamefont {D.~Y.}\ \bibnamefont {Choi}},
  \bibinfo {author} {\bibfnamefont {I.}~\bibnamefont {Staude}}, \bibinfo
  {author} {\bibfnamefont {A.~E.}\ \bibnamefont {Miroshnichenko}}, \bibinfo
  {author} {\bibfnamefont {D.~N.}\ \bibnamefont {Neshev}}, \bibinfo {author}
  {\bibfnamefont {A.~A.}\ \bibnamefont {Fedyanin}}, \ and\ \bibinfo {author}
  {\bibfnamefont {Y.~S.}\ \bibnamefont {Kivshar}},\ }\href {\doibase
  10.1021/acs.nanolett.5b02989} {\bibfield  {journal} {\bibinfo  {journal}
  {Nano Lett.}\ }\textbf {\bibinfo {volume} {15}},\ \bibinfo {pages} {6985}
  (\bibinfo {year} {2015})}\BibitemShut {NoStop}%
\bibitem [{\citenamefont {Makarov}\ \emph {et~al.}(2015)\citenamefont
  {Makarov}, \citenamefont {Kudryashov}, \citenamefont {Mukhin}, \citenamefont
  {Mozharov}, \citenamefont {Milichko}, \citenamefont {Krasnok},\ and\
  \citenamefont {Belov}}]{Makarov2015}%
  \BibitemOpen
  \bibfield  {author} {\bibinfo {author} {\bibfnamefont {S.}~\bibnamefont
  {Makarov}}, \bibinfo {author} {\bibfnamefont {S.}~\bibnamefont {Kudryashov}},
  \bibinfo {author} {\bibfnamefont {I.}~\bibnamefont {Mukhin}}, \bibinfo
  {author} {\bibfnamefont {A.}~\bibnamefont {Mozharov}}, \bibinfo {author}
  {\bibfnamefont {V.}~\bibnamefont {Milichko}}, \bibinfo {author}
  {\bibfnamefont {A.}~\bibnamefont {Krasnok}}, \ and\ \bibinfo {author}
  {\bibfnamefont {P.}~\bibnamefont {Belov}},\ }\href@noop {} {\bibfield
  {journal} {\bibinfo  {journal} {Nano Lett.}\ }\textbf {\bibinfo {volume}
  {15}},\ \bibinfo {pages} {6187} (\bibinfo {year} {2015})}\BibitemShut
  {NoStop}%
\bibitem [{\citenamefont {Baranov}\ \emph
  {et~al.}(2016{\natexlab{a}})\citenamefont {Baranov}, \citenamefont {Makarov},
  \citenamefont {Milichko}, \citenamefont {Kudryashov}, \citenamefont
  {Krasnok},\ and\ \citenamefont {Belov}}]{Baranov2016}%
  \BibitemOpen
  \bibfield  {author} {\bibinfo {author} {\bibfnamefont {D.~G.}\ \bibnamefont
  {Baranov}}, \bibinfo {author} {\bibfnamefont {S.~V.}\ \bibnamefont
  {Makarov}}, \bibinfo {author} {\bibfnamefont {V.~A.}\ \bibnamefont
  {Milichko}}, \bibinfo {author} {\bibfnamefont {S.~I.}\ \bibnamefont
  {Kudryashov}}, \bibinfo {author} {\bibfnamefont {A.~E.}\ \bibnamefont
  {Krasnok}}, \ and\ \bibinfo {author} {\bibfnamefont {P.~A.}\ \bibnamefont
  {Belov}},\ }\href {\doibase 10.1021/acsphotonics.6b00358} {\bibfield
  {journal} {\bibinfo  {journal} {ACS Photonics}\ }\textbf {\bibinfo {volume}
  {3}},\ \bibinfo {pages} {1546} (\bibinfo {year}
  {2016}{\natexlab{a}})}\BibitemShut {NoStop}%
\bibitem [{\citenamefont {Fischer}\ \emph {et~al.}(2016)\citenamefont
  {Fischer}, \citenamefont {Schmidt}, \citenamefont {Sakat}, \citenamefont
  {Stock}, \citenamefont {Samarelli}, \citenamefont {Frigerio}, \citenamefont
  {Ortolani}, \citenamefont {Paul}, \citenamefont {Isella}, \citenamefont
  {Leitenstorfer}, \citenamefont {Biagioni},\ and\ \citenamefont
  {Brida}}]{Fischer2016}%
  \BibitemOpen
  \bibfield  {author} {\bibinfo {author} {\bibfnamefont {M.~P.}\ \bibnamefont
  {Fischer}}, \bibinfo {author} {\bibfnamefont {C.}~\bibnamefont {Schmidt}},
  \bibinfo {author} {\bibfnamefont {E.}~\bibnamefont {Sakat}}, \bibinfo
  {author} {\bibfnamefont {J.}~\bibnamefont {Stock}}, \bibinfo {author}
  {\bibfnamefont {A.}~\bibnamefont {Samarelli}}, \bibinfo {author}
  {\bibfnamefont {J.}~\bibnamefont {Frigerio}}, \bibinfo {author}
  {\bibfnamefont {M.}~\bibnamefont {Ortolani}}, \bibinfo {author}
  {\bibfnamefont {D.~J.}\ \bibnamefont {Paul}}, \bibinfo {author}
  {\bibfnamefont {G.}~\bibnamefont {Isella}}, \bibinfo {author} {\bibfnamefont
  {A.}~\bibnamefont {Leitenstorfer}}, \bibinfo {author} {\bibfnamefont
  {P.}~\bibnamefont {Biagioni}}, \ and\ \bibinfo {author} {\bibfnamefont
  {D.}~\bibnamefont {Brida}},\ }\href {\doibase 10.1103/PhysRevLett.117.047401}
  {\bibfield  {journal} {\bibinfo  {journal} {Phys. Rev. Lett.}\ }\textbf
  {\bibinfo {volume} {117}},\ \bibinfo {pages} {047401} (\bibinfo {year}
  {2016})}\BibitemShut {NoStop}%
\bibitem [{\citenamefont {Baranov}\ \emph
  {et~al.}(2016{\natexlab{b}})\citenamefont {Baranov}, \citenamefont {Makarov},
  \citenamefont {Krasnok}, \citenamefont {Belov},\ and\ \citenamefont
  {Al\`u}}]{BaranovLPR}%
  \BibitemOpen
  \bibfield  {author} {\bibinfo {author} {\bibfnamefont {D.~G.}\ \bibnamefont
  {Baranov}}, \bibinfo {author} {\bibfnamefont {S.~V.}\ \bibnamefont
  {Makarov}}, \bibinfo {author} {\bibfnamefont {A.~E.}\ \bibnamefont
  {Krasnok}}, \bibinfo {author} {\bibfnamefont {P.~A.}\ \bibnamefont {Belov}},
  \ and\ \bibinfo {author} {\bibfnamefont {A.}~\bibnamefont {Al\`u}},\
  }\href@noop {} {\bibfield  {journal} {\bibinfo  {journal} {Laser Photon.
  Rev.}\ }\textbf {\bibinfo {volume} {10}},\ \bibinfo {pages} {1009} (\bibinfo
  {year} {2016}{\natexlab{b}})}\BibitemShut {NoStop}%
\bibitem [{\citenamefont {Rauschera}\ and\ \citenamefont
  {Laenen}(1997)}]{GeTPA}%
  \BibitemOpen
  \bibfield  {author} {\bibinfo {author} {\bibfnamefont {C.}~\bibnamefont
  {Rauschera}}\ and\ \bibinfo {author} {\bibfnamefont {R.}~\bibnamefont
  {Laenen}},\ }\href@noop {} {\bibfield  {journal} {\bibinfo  {journal} {J.
  Appl. Phys.}\ }\textbf {\bibinfo {volume} {81}},\ \bibinfo {pages} {2818}
  (\bibinfo {year} {1997})}\BibitemShut {NoStop}%
\bibitem [{\citenamefont {Krasnok}\ \emph {et~al.}(2016)\citenamefont
  {Krasnok}, \citenamefont {Glybovski}, \citenamefont {Petrov}, \citenamefont
  {Makarov}, \citenamefont {Savelev}, \citenamefont {Belov}, \citenamefont
  {Simovski},\ and\ \citenamefont {Kivshar}}]{KrasnokAPL2016}%
  \BibitemOpen
  \bibfield  {author} {\bibinfo {author} {\bibfnamefont {A.~E.}\ \bibnamefont
  {Krasnok}}, \bibinfo {author} {\bibfnamefont {S.~B.}\ \bibnamefont
  {Glybovski}}, \bibinfo {author} {\bibfnamefont {M.~I.}\ \bibnamefont
  {Petrov}}, \bibinfo {author} {\bibfnamefont {S.~V.}\ \bibnamefont {Makarov}},
  \bibinfo {author} {\bibfnamefont {R.~S.}\ \bibnamefont {Savelev}}, \bibinfo
  {author} {\bibfnamefont {P.~A.}\ \bibnamefont {Belov}}, \bibinfo {author}
  {\bibfnamefont {C.~R.}\ \bibnamefont {Simovski}}, \ and\ \bibinfo {author}
  {\bibfnamefont {Y.~S.}\ \bibnamefont {Kivshar}},\ }\href@noop {} {\bibfield
  {journal} {\bibinfo  {journal} {Appl. Phys. Lett.}\ }\textbf {\bibinfo
  {volume} {108}},\ \bibinfo {pages} {211105} (\bibinfo {year}
  {2016})}\BibitemShut {NoStop}%
\bibitem [{\citenamefont {Shore}\ and\ \citenamefont
  {Yaghjian}(2012{\natexlab{a}})}]{ShoreRS2012}%
  \BibitemOpen
  \bibfield  {author} {\bibinfo {author} {\bibfnamefont {R.~A.}\ \bibnamefont
  {Shore}}\ and\ \bibinfo {author} {\bibfnamefont {A.~D.}\ \bibnamefont
  {Yaghjian}},\ }\href@noop {} {\bibfield  {journal} {\bibinfo  {journal}
  {Radio Science}\ }\textbf {\bibinfo {volume} {47}} (\bibinfo {year}
  {2012}{\natexlab{a}})}\BibitemShut {NoStop}%
\bibitem [{\citenamefont {Huldt}(1974)}]{GeAuger}%
  \BibitemOpen
  \bibfield  {author} {\bibinfo {author} {\bibfnamefont {L.}~\bibnamefont
  {Huldt}},\ }\href@noop {} {\bibfield  {journal} {\bibinfo  {journal} {Phys.
  Stat. Sol.}\ }\textbf {\bibinfo {volume} {24}},\ \bibinfo {pages} {221}
  (\bibinfo {year} {1974})}\BibitemShut {NoStop}%
\bibitem [{\citenamefont {Wurtz}\ \emph {et~al.}(2011)\citenamefont {Wurtz},
  \citenamefont {Pollard}, \citenamefont {Hendren}, \citenamefont
  {Wiederrecht}, \citenamefont {Gosztola}, \citenamefont {Podolskiy},\ and\
  \citenamefont {Zayats}}]{Wurtz}%
  \BibitemOpen
  \bibfield  {author} {\bibinfo {author} {\bibfnamefont {G.~A.}\ \bibnamefont
  {Wurtz}}, \bibinfo {author} {\bibfnamefont {R.}~\bibnamefont {Pollard}},
  \bibinfo {author} {\bibfnamefont {W.}~\bibnamefont {Hendren}}, \bibinfo
  {author} {\bibfnamefont {G.~P.}\ \bibnamefont {Wiederrecht}}, \bibinfo
  {author} {\bibfnamefont {D.~J.}\ \bibnamefont {Gosztola}}, \bibinfo {author}
  {\bibfnamefont {V.~A.}\ \bibnamefont {Podolskiy}}, \ and\ \bibinfo {author}
  {\bibfnamefont {A.~V.}\ \bibnamefont {Zayats}},\ }\href@noop {} {\bibfield
  {journal} {\bibinfo  {journal} {Nat. Nanotechnol.}\ }\textbf {\bibinfo
  {volume} {6}},\ \bibinfo {pages} {107} (\bibinfo {year} {2011})}\BibitemShut
  {NoStop}%
\bibitem [{\citenamefont {Valev}\ \emph {et~al.}(2012)\citenamefont {Valev},
  \citenamefont {Denkova}, \citenamefont {Zheng}, \citenamefont {Kuznetsov},
  \citenamefont {Reinhardt}, \citenamefont {Chichkov}, \citenamefont
  {Tsutsumanova}, \citenamefont {Osley}, \citenamefont {Petkov}, \citenamefont
  {De~Clercq}, \citenamefont {Silhanek}, \citenamefont {Jeyaram}, \citenamefont
  {Volskiy}, \citenamefont {Warburton}, \citenamefont {Vandenbosch},
  \citenamefont {Russev}, \citenamefont {Aktsipetrov}, \citenamefont {Ameloot},
  \citenamefont {Moshchalkov},\ and\ \citenamefont {Verbiest}}]{Valev}%
  \BibitemOpen
  \bibfield  {author} {\bibinfo {author} {\bibfnamefont {V.~K.}\ \bibnamefont
  {Valev}}, \bibinfo {author} {\bibfnamefont {D.}~\bibnamefont {Denkova}},
  \bibinfo {author} {\bibfnamefont {X.}~\bibnamefont {Zheng}}, \bibinfo
  {author} {\bibfnamefont {A.~I.}\ \bibnamefont {Kuznetsov}}, \bibinfo {author}
  {\bibfnamefont {C.}~\bibnamefont {Reinhardt}}, \bibinfo {author}
  {\bibfnamefont {B.~N.}\ \bibnamefont {Chichkov}}, \bibinfo {author}
  {\bibfnamefont {G.}~\bibnamefont {Tsutsumanova}}, \bibinfo {author}
  {\bibfnamefont {E.~J.}\ \bibnamefont {Osley}}, \bibinfo {author}
  {\bibfnamefont {V.}~\bibnamefont {Petkov}}, \bibinfo {author} {\bibfnamefont
  {B.}~\bibnamefont {De~Clercq}}, \bibinfo {author} {\bibfnamefont {A.~V.}\
  \bibnamefont {Silhanek}}, \bibinfo {author} {\bibfnamefont {Y.}~\bibnamefont
  {Jeyaram}}, \bibinfo {author} {\bibfnamefont {V.}~\bibnamefont {Volskiy}},
  \bibinfo {author} {\bibfnamefont {P.~A.}\ \bibnamefont {Warburton}}, \bibinfo
  {author} {\bibfnamefont {G.~A.~E.}\ \bibnamefont {Vandenbosch}}, \bibinfo
  {author} {\bibfnamefont {S.}~\bibnamefont {Russev}}, \bibinfo {author}
  {\bibfnamefont {O.~A.}\ \bibnamefont {Aktsipetrov}}, \bibinfo {author}
  {\bibfnamefont {M.}~\bibnamefont {Ameloot}}, \bibinfo {author} {\bibfnamefont
  {V.~V.}\ \bibnamefont {Moshchalkov}}, \ and\ \bibinfo {author} {\bibfnamefont
  {T.}~\bibnamefont {Verbiest}},\ }\href {\doibase 10.1002/adma.201103807}
  {\bibfield  {journal} {\bibinfo  {journal} {Adv. Mater.}\ }\textbf {\bibinfo
  {volume} {24}},\ \bibinfo {pages} {OP29} (\bibinfo {year}
  {2012})}\BibitemShut {NoStop}%
\bibitem [{\citenamefont {Mulholland}\ \emph {et~al.}(1994)\citenamefont
  {Mulholland}, \citenamefont {Bohren},\ and\ \citenamefont
  {Fuller}}]{MulhollandLangmuir1994}%
  \BibitemOpen
  \bibfield  {author} {\bibinfo {author} {\bibfnamefont {G.~W.}\ \bibnamefont
  {Mulholland}}, \bibinfo {author} {\bibfnamefont {C.~F.}\ \bibnamefont
  {Bohren}}, \ and\ \bibinfo {author} {\bibfnamefont {K.~A.}\ \bibnamefont
  {Fuller}},\ }\href@noop {} {\bibfield  {journal} {\bibinfo  {journal}
  {Langmuir}\ }\textbf {\bibinfo {volume} {10}},\ \bibinfo {pages} {2533}
  (\bibinfo {year} {1994})}\BibitemShut {NoStop}%
\bibitem [{\citenamefont {Merchiers}\ \emph {et~al.}(2007)\citenamefont
  {Merchiers}, \citenamefont {Moreno}, \citenamefont {Gonz\'alez},\ and\
  \citenamefont {Saiz}}]{MerchiersPRA2007}%
  \BibitemOpen
  \bibfield  {author} {\bibinfo {author} {\bibfnamefont {O.}~\bibnamefont
  {Merchiers}}, \bibinfo {author} {\bibfnamefont {F.}~\bibnamefont {Moreno}},
  \bibinfo {author} {\bibfnamefont {F.}~\bibnamefont {Gonz\'alez}}, \ and\
  \bibinfo {author} {\bibfnamefont {J.~M.}\ \bibnamefont {Saiz}},\ }\href@noop
  {} {\bibfield  {journal} {\bibinfo  {journal} {Phys. Rev. A}\ }\textbf
  {\bibinfo {volume} {76}},\ \bibinfo {pages} {043834} (\bibinfo {year}
  {2007})}\BibitemShut {NoStop}%
\bibitem [{\citenamefont {Savelev}\ \emph {et~al.}(2014)\citenamefont
  {Savelev}, \citenamefont {Slobozhanyuk}, \citenamefont {Miroshnichenko},
  \citenamefont {Kivshar},\ and\ \citenamefont {Belov}}]{SavelevPRB2014}%
  \BibitemOpen
  \bibfield  {author} {\bibinfo {author} {\bibfnamefont {R.}~\bibnamefont
  {Savelev}}, \bibinfo {author} {\bibfnamefont {A.}~\bibnamefont
  {Slobozhanyuk}}, \bibinfo {author} {\bibfnamefont {A.}~\bibnamefont
  {Miroshnichenko}}, \bibinfo {author} {\bibfnamefont {Y.}~\bibnamefont
  {Kivshar}}, \ and\ \bibinfo {author} {\bibfnamefont {P.}~\bibnamefont
  {Belov}},\ }\href@noop {} {\bibfield  {journal} {\bibinfo  {journal} {Phys.
  Rev. B}\ }\textbf {\bibinfo {volume} {89}},\ \bibinfo {pages} {035435}
  (\bibinfo {year} {2014})}\BibitemShut {NoStop}%
\bibitem [{\citenamefont {Shore}\ and\ \citenamefont
  {Yaghjian}(2012{\natexlab{b}})}]{ShoreRS2012part1}%
  \BibitemOpen
  \bibfield  {author} {\bibinfo {author} {\bibfnamefont {R.~A.}\ \bibnamefont
  {Shore}}\ and\ \bibinfo {author} {\bibfnamefont {A.~D.}\ \bibnamefont
  {Yaghjian}},\ }\href@noop {} {\bibfield  {journal} {\bibinfo  {journal}
  {Radio Science}\ }\textbf {\bibinfo {volume} {47}} (\bibinfo {year}
  {2012}{\natexlab{b}})}\BibitemShut {NoStop}%
\bibitem [{\citenamefont {Novotny}\ and\ \citenamefont {van
  Hulst}(2011{\natexlab{b}})}]{NovotnyNP2010}%
  \BibitemOpen
  \bibfield  {author} {\bibinfo {author} {\bibfnamefont {L.}~\bibnamefont
  {Novotny}}\ and\ \bibinfo {author} {\bibfnamefont {N.}~\bibnamefont {van
  Hulst}},\ }\href@noop {} {\bibfield  {journal} {\bibinfo  {journal} {Nature
  Photonics}\ }\textbf {\bibinfo {volume} {5}},\ \bibinfo {pages} {83}
  (\bibinfo {year} {2011}{\natexlab{b}})}\BibitemShut {NoStop}%
\bibitem [{\citenamefont {Novotny}\ and\ \citenamefont
  {Hecht}(2006)}]{NovotnyPrinciplesNanoOptics}%
  \BibitemOpen
  \bibfield  {author} {\bibinfo {author} {\bibfnamefont {L.}~\bibnamefont
  {Novotny}}\ and\ \bibinfo {author} {\bibfnamefont {B.}~\bibnamefont
  {Hecht}},\ }\href@noop {} {\emph {\bibinfo {title} {Principles of
  Nano-Optics}}}\ (\bibinfo  {publisher} {Cambridge University Press},\
  \bibinfo {year} {2006})\BibitemShut {NoStop}%
\bibitem [{\citenamefont {Al\`u}\ and\ \citenamefont
  {Engheta}(2006)}]{AluPRB2006}%
  \BibitemOpen
  \bibfield  {author} {\bibinfo {author} {\bibfnamefont {A.}~\bibnamefont
  {Al\`u}}\ and\ \bibinfo {author} {\bibfnamefont {N.}~\bibnamefont
  {Engheta}},\ }\href@noop {} {\bibfield  {journal} {\bibinfo  {journal} {Phys.
  Rev. B}\ }\textbf {\bibinfo {volume} {74}},\ \bibinfo {pages} {205436}
  (\bibinfo {year} {2006})}\BibitemShut {NoStop}%
\bibitem [{\citenamefont {Weber}\ and\ \citenamefont
  {Ford}(2004)}]{WeberPRB2004}%
  \BibitemOpen
  \bibfield  {author} {\bibinfo {author} {\bibfnamefont {W.~H.}\ \bibnamefont
  {Weber}}\ and\ \bibinfo {author} {\bibfnamefont {G.~W.}\ \bibnamefont
  {Ford}},\ }\href@noop {} {\bibfield  {journal} {\bibinfo  {journal} {Phys.
  Rev. B}\ }\textbf {\bibinfo {volume} {70}},\ \bibinfo {pages} {125429}
  (\bibinfo {year} {2004})}\BibitemShut {NoStop}%
\end{thebibliography}

\end{document}